\documentclass[superscriptaddress,twocolumn,showpacs,a4paper,
amssymb,amsmath,nobibnotes,aps,prd,
showkeys,
nofootinbib,notitlepage]{revtex4-1}
\usepackage{verbatim}
\usepackage[T1]{fontenc}
\usepackage[utf8]{inputenc}
\usepackage[american]{babel}
\usepackage{epsfig}
\usepackage{booktabs}
\usepackage{multirow}
\usepackage{dcolumn}
\usepackage{amsmath}
\usepackage{mathtools}
\usepackage{amsfonts}
\usepackage{amssymb}
\usepackage{ulem}
\usepackage{epstopdf}
\usepackage{bm}
\usepackage{siunitx}
\usepackage{braket}
\usepackage{enumitem}
\usepackage{soul}
\usepackage[table]{xcolor}
\usepackage{color}
\usepackage{transparent}
\usepackage{pifont}
\usepackage{enumitem}

\definecolor{navyblue}{rgb}{0.0, 0.0, 0.5}
\definecolor{royalblue}{rgb}{0.25, 0.41, 0.88}
\definecolor{cadmiumgreen}{rgb}{0.0, 0.42, 0.24}
\definecolor{blue-violet}{rgb}{0.54, 0.17, 0.89}
\definecolor{darkviolet}{rgb}{0.58, 0.0, 0.83}
\definecolor{orange(colorwheel)}{rgb}{1.0, 0.5, 0.0}

\usepackage{hyperref}
\hypersetup{
    colorlinks=true, 
    linkcolor=royalblue, 
    citecolor=magenta}

\usepackage{booktabs}
\usepackage{multirow}
\usepackage{dcolumn}
\usepackage{colortbl}

\newcommand{\nq}[1]{%
	\begin{tabular}{@{}c@{}}\strut#1\strut\end{tabular}%
}

\begin{document}

\title{Hints of Neutrino Dark Matter scattering in the CMB? Constraints from the Marginalized and Profile Distributions}

\author{William Giar\`{e}}
\email{w.giare@sheffield.ac.uk}
\affiliation{Consortium for Fundamental Physics, School of Mathematics and Statistics,
University of Sheffield, Hounsfield Road, Sheffield S3 7RH, United Kingdom}

\author{Adri\`a G\'omez-Valent}
\email{agomezvalent@icc.ub.edu}
\affiliation{Departament de Física Quàntica i Astrofísica, and Institute of Cosmos Sciences, Universitat de Barcelona,
Av. Diagonal 647, 08028 Barcelona, Spain}

\author{Eleonora Di Valentino}
\email{e.divalentino@sheffield.ac.uk}
\affiliation{Consortium for Fundamental Physics, School of Mathematics and Statistics,
University of Sheffield, Hounsfield Road, Sheffield S3 7RH, United Kingdom}

\author{Carsten van de Bruck}
\email{c.vandebruck@sheffield.ac.uk}
\affiliation{Consortium for Fundamental Physics, School of Mathematics and Statistics,
University of Sheffield, Hounsfield Road, Sheffield S3 7RH, United Kingdom}

\begin{abstract}
\noindent We study scatter-like interactions between neutrinos and dark matter in light of different combinations of temperature, polarization and lensing data released by three independent CMB experiments -- the Planck satellite, the Atacama Cosmology Telescope (ACT), and the South Pole Telescope (SPT) -- in conjunction with Baryon Acoustic Oscillation (BAO) measurements. We apply two different statistical methodologies. Alongside the usual marginalization technique, we cross-check all the results through a Profile Likelihood analysis. As a first step, working under the assumption of massless neutrinos, we perform a comprehensive (re-)analysis aimed at assessing the validity of some recent results hinting at a mild preference for non-vanishing interactions from small-scale CMB data. We find compelling resilience in the results already documented in the literature, confirming that interactions with a strength $u_{\nu\rm{DM}} \sim 10^{-5} - 10^{-4}$ appear to be globally favored by ACT (both alone and in combination with Planck). This result is corroborated by the inclusion of additional data, such as the recent ACT-DR6 lensing likelihood, as well as by the Profile Likelihood analysis. Interestingly, a fully consistent preference for interactions emerges from SPT, as well, although it is weaker than the one obtained from ACT. As a second step, we repeat the same analysis considering neutrinos as massive particles. Despite the larger parameter space, all the hints pointing towards interactions are confirmed also in this more realistic case. In addition, we report a very mild preference for interactions in Planck+BAO alone (not found in the massless case) which aligns with small-scale data. While this latter result is not fully confirmed by the Profile Likelihood analysis, the profile distribution does confirm that interactions are not disfavoured by Planck. 
\end{abstract}

\keywords{cosmological parameters -- Cosmology: observations -- Cosmology: theory -- dark matter}

\maketitle

\section{Introduction}\label{sec:intro}

In recent decades, remarkable advancements in the field of observational cosmology and astrophysics have provided a refined tool for testing our understanding of fundamental physics, revealing evidence (both direct and indirect) of previously unknown and unforeseen phenomena. 

In the list of groundbreaking discoveries, we can mention the fact that the current expansion of the Universe is accelerating (as originally determined via observations of distant Type Ia Supernovae~\cite{SupernovaSearchTeam:1998fmf,SupernovaCosmologyProject:1998vns}, and now directly and indirectly corroborated by a wide variety of other probes~\cite{Sherwin:2011gv,Moresco:2016mzx,Haridasu:2017lma,Rubin:2016iqe,Planck:2018vyg,Gomez-Valent:2018gvm,Yang:2019fjt,Nadathur:2020kvq,Rose:2020shp,DiValentino:2020evt,eBOSS:2020yzd,Escamilla:2023oce}) and the indirect evidence for a non-baryonic matter, commonly known as Dark Matter (DM), that was originally highlighted through anomalies in the rotation of galaxies and the velocities of stars within them~\cite{Kapteyn:1922zz,Zwicky:1933gu,Zwicky:1937zza,Freeman:1970mx,Rubin:1970zza}, and subsequently supported by a multitude of other observations, including measurements of the Cosmic Microwave Background (CMB) and the large-scale structure of the Universe~\cite{WMAP:2012fli,WMAP:2012nax,Planck:2018nkj,Planck:2018vyg,SPT:2004qip,SPT-3G:2021eoc,ACT:2020frw,ACT:2020gnv,ACT:2023kun,BOSS:2012dmf,BOSS:2013rlg,BOSS:2014hwf,BOSS:2016wmc,BOSS:2013uda,eBOSS:2020yzd,SDSS:2003eyi,SDSS:2004kqt,SDSS:2006lmn,SDSS:2014iwm,DES:2016jjg,DES:2017qwj,DES:2021wwk,DES:2022ccp,KiDS:2020suj,deJong:2012zb,KiDS:2020ghu,Kilo-DegreeSurvey:2023gfr,Pan-STARRS1:2017jku,Brout:2022vxf}. 

These discoveries have significantly advanced our understanding of the Universe, leading to the development of the standard $\Lambda$CDM model of cosmology which provides a robust framework to describe observational data probing very different cosmic epochs and scales. However, despite this remarkable success, it is not an exaggeration to say that our comprehension of the ingredients of the model, such as DM and Dark Energy (DE), remains largely elusive, lacking a well-established theoretical foundation as well as direct experimental confirmation.

The most stringent constraints on the properties of DM arise from observations in the field of cosmology and ongoing efforts in direct detection experiments~\cite{Roszkowski:2017nbc,Billard:2021uyg,Buen-Abad:2021mvc}. Despite several attempts aimed at explaining DM in terms of modified gravity, the prevailing consensus is that it consists of non-relativistic (or `cold') massive particles that do not interact with other particles, except through gravity. Within this interpretation, the enigmatic nature of DM is actually rooted in its poorly understood interactions with other particles that remain a subject of ongoing debate and study. Several possible channels of interactions have been tested and studied both in cosmology and particle physics, including interactions with photons, baryons, dark radiations and neutrinos,\footnote{More exotic scenarios that involve DM-DE interactions have been largely considered in light of their apparent ability to reduce current tensions in cosmology, see e.g., Refs.~\cite{Mangano:2006mp,Valiviita:2008iv,Gavela:2009cy,Salvatelli:2014zta,Sola:2016jky,DiValentino:2017iww, Kumar:2017dnp, Wang:2016lxa,SolaPeracaula:2017esw,Sola:2017znb,Gomez-Valent:2018nib, Martinelli:2019dau, Yang:2019uog, DiValentino:2019ffd, Pan:2019jqh, Kumar:2019wfs, Yang:2018euj,Escudero:2015yka, Kumar:2016zpg, Murgia:2016ccp,Pourtsidou:2016ico, Yang:2018ubt, Barros:2018efl,Yang:2019uzo, Pan:2019gop, DiValentino:2019jae, Stadler:2019dii, DiValentino:2020leo, Yao:2020pji, Lucca:2020zjb, DiValentino:2020kpf, Gomez-Valent:2020mqn, Yang:2020uga, Yao:2020hkw, Pan:2020bur, DiValentino:2020vnx, Mosbech:2020ahp,Hogg:2020rdp, SolaPeracaula:2021gxi,Lucca:2021dxo, Kumar:2021eev, Yang:2021hxg, Gao:2021xnk, Yang:2021oxc, Lucca:2021eqy, Halder:2021jiv, Kaneta:2022kjj, Gariazzo:2021qtg, Nunes:2021zzi, Yang:2022csz, Nunes:2022bhn,Goh:2022gxo,Gomez-Valent:2022bku,Barros:2022bdv,vanderWesthuizen:2023hcl,Zhai:2023yny,Bernui:2023byc,Oikonomou:2023qfl,deCruzPerez:2023wzd,SolaPeracaula:2023swx} and Refs.~\cite{DiValentino:2021izs,Knox:2019rjx,Jedamzik:2020zmd,Abdalla:2022yfr,Kamionkowski:2022pkx} for more general reviews. } see, e.g., Refs.~\cite{Cyr-Racine:2015ihg, Becker:2020hzj,Audren:2014bca,Lesgourgues:2015wza,Archidiacono:2019wdp,Dvorkin:2013cea,Munoz:2015bca,Slatyer:2018aqg,Xu:2018efh,Stadler:2018jin,Palomares-Ruiz:2007trf,Serra:2009uu,Shoemaker:2013tda,Wilkinson:2013kia,Wilkinson:2014ksa,Olivares-DelCampo:2017feq,Bertoni:2014mva,Batell:2017rol,Batell:2017cmf,DiValentino:2017oaw,Escudero:2018thh,Kolb:1987qy,Shoemaker:2015qul,deSalas:2016svi,Pandey:2018wvh,Kelly:2019wow,Blennow:2019fhy,Choi:2019ixb,RoyChoudhury:2020dmd,Kelly:2021mcd,RoyChoudhury:2022rva,Mosbech:2022nkk,Lin:2023nsm,Akita:2023yga,Bostan:2023ped}. 

Despite the vast majority of these studies not yielding any conclusive evidence supporting DM interactions, recently, a few scattered and very mild hints in favor of an elastic scattering between neutrinos and DM ($\nu$DM) have been documented in the literature. Originally, in Ref.~\cite{Hooper:2021rjc} a preference for $\nu$DM interactions at the level of $\sim 3\sigma$ has been found by analyzing Lyman-$\alpha$ data. Subsequently, in Ref.~\cite{Brax:2023rrf} it was argued that observations of temperature and polarization anisotropies in the CMB at small angular scales could be crucial in revealing unique signatures for $\nu$DM interactions that would be challenging to detect on larger angular scales. This has motivated a comprehensive re-analysis of $\nu$DM interactions in light of the most recent small-scale CMB data and in Refs.~\cite{Brax:2023rrf, Brax:2023tvn} a very mild preference for interactions (in perfect agreement with Lyman-$\alpha$ data) emerged at a statistical significance ranging between one and two standard deviations.  

Needless to say, taking each of these individual results at face value, \textit{none} of them is enough to claim any compelling evidence for $\nu$DM interactions. However, while exercising caution is a fundamental requirement, it is equally imperative to acknowledge that these hints present intriguing signals that at least warrant further investigation and rigorous cross-checking. This holds particularly true in the context of next-generation CMB experiments whose declared goal is to significantly enhance our understanding of neutrino physics and the dark sector of the cosmological model.

In this paper, we take other steps forward in this direction and carry out a comprehensive re-analysis of these scattered hints aimed at definitively assessing their robustness and validity. More specifically, we broaden our analysis of the CMB data to address the following key aspects:
\begin{itemize}

\item We extend the study of $\nu$DM interactions to \textit{all} the most recent CMB temperature, polarization and lensing measurements, as well as to different combinations of them. This includes novel and independent measurements, such as the recent lensing Data Release 6 from the Atacama Cosmology Telescope (ACT)~\cite{ACT:2023kun,ACT:2023dou} that has provided a comprehensive reconstruction of CMB lensing over 9400 sq. deg. of the sky\footnote{Precise reconstructions of the DM distribution through the lensing spectrum hold particular relevance in the study of the properties of neutrinos and other light particles, see, e.g., Ref.~\cite{Giare:2023aix}, where late-time-only constraints on the total neutrino mass were derived and are often stronger than early-time constraints in certain extended cosmologies~\cite{RoyChoudhury:2018gay,RoyChoudhury:2019hls,DiValentino:2021imh,DiValentino:2022oon,DiValentino:2022njd,Forconi:2023akg}. } and the measurements of temperature and polarization anisotropies released by the South Pole Telescope (SPT)~\cite{SPT-3G:2022hvq,SPT-3G:2021eoc}. 

\item For \textit{all} experiments, we re-analyze $\nu$DM interactions under the assumption of massless neutrinos. Then, we address this theoretical approximation and extend the results to $\nu$DM interactions in the presence of massive neutrinos. To maintain generality, we treat the mass value as a free parameter to be determined by data.

\item We asses the validity of \textit{all} our findings using the commonly used Markov Chain Monte Carlo (MCMC) method and applying two different (complementary) statistical methodologies to interpret the information stored in the chains. Alongside the usual marginalization technique, we cross-check all our results through a Profile Likelihood analysis in order to duly assess the impact of volume effects on our results.

\end{itemize}

Our findings reveal a remarkable robustness of these trends: despite the statistical significance of our results may be limited given the current data sensitivity, their resilience becomes apparent through cross-validation in independent experiments, as well as through the consistency observed when comparing the marginalized and profile distributions.

The paper is organized as follows: in \autoref{sec:theory}, we briefly review the theoretical framework adopted for describing $\nu$DM interactions, as well as their cosmological phenomenology. In \autoref{sec:method}, we discuss the methodology and data exploited in this work. In \autoref{sec:results_massless}, we systematically derive and discuss the constraints on $\nu$DM interactions from all available CMB experiments (and combinations of them) under the assumption of massless neutrinos. In \autoref{sec:results_massive}, we repeat the same analyses, considering neutrinos as massive particles whose mass value has to be determined by data. In both these sections we cross-check all our findings using both marginalization and Profile Likelihood analyses. Finally, in \autoref{sec:conclusions}, we summarize our major conclusions.

\section{\boldmath{$\nu$}DM Interactions}\label{sec:theory}
We provide a concise overview of the theoretical framework adopted for describing $\nu$DM interactions, along with a brief discussion of their cosmological implications. In this section our goal is to provide some background useful to correctly interpret and clarify the results presented later on in the manuscript. 

We work in the Newtonian gauge characterized by  two scalar potentials $\psi$ and $\phi$ which appear in the line element of a perturbed flat Friedmann-Lema\^itre-Robertson-Walker universe as 
\begin{equation}
d s^2=a^2(\tau)\left[-(1+2 \psi) d \tau^2+(1-2 \phi) d x_i d x^i\right],
\label{eq:gauge}
\end{equation}
where $d\tau=dt/a(\tau)$ is the conformal time and $x^i$ the comoving coordinates. 

In order to derive equations for perturbations, we start from the relativistic Boltzmann equation~\cite{Ma:1995ey,Oldengott:2014qra}:

\begin{equation}
   \frac{\partial f}{\partial \tau}+\frac{dx^i}{d\tau}\frac{\partial f}{\partial x^i} +\frac{dp}{d\tau}\frac{\partial f}{\partial p}+\frac{dn_i}{d\tau}\frac{\partial f}{\partial n_i}=\left(\frac{\partial f}{\partial \tau}\right)_C\,,
\end{equation}
with $n_i$ the components of  a unit vector \boldsymbol{$\hat n$}  pointing in the direction of the comoving momentum $\mathbf{p}$, i.e. $\mathbf{p}=p\hat{\mathbf{n}}$, and $f$ the distribution function, which is defined through the number of particles in a phase-space volume element
\begin{equation}
f(x^i, P_j, \tau) d x^1 d x^2 d x^3 d P_1 d P_2 d P_3=d N,
\end{equation}
being $P_i$ the canonical momentum conjugate. Defining the perturbation of the distribution function, $\Psi$, as 
\begin{equation}
f(x^i, P_j, \tau)=f^{(0)}(p)[1+\Psi(\mathbf{x}, \mathbf{p}, \tau)],
\end{equation}
in the Newtonian gauge, the Boltzmann equation for $\Psi$ in momentum space reduces to~\cite{Ma:1995ey}
\begin{equation}
\begin{aligned}
\frac{\partial \Psi}{\partial \tau}+i \frac{p}{E}(\mathbf{k} \cdot \hat{\mathbf{n}}) \Psi+\frac{d \ln f^{(0)}(p)}{d \ln p}\left[\dot{\phi}-i \frac{E}{p}(\mathbf{k} \cdot \hat{\mathbf{n}}) \psi\right]& \\= \frac{1}{f^{(0)}(p) }\left(\frac{\partial f}{\partial \tau}\right)_C,
\label{eq:Boltzmann}
\end{aligned}
\end{equation}
where dots denote the derivatives with respect to conformal time and $E$ is the comoving particle energy. 

\subsection{\boldmath{${\nu}$}DM Perturbations for massless neutrinos}
\label{sec.theory.massless}
We start deriving the equations for perturbations under the assumption of massless neutrinos. Starting from Eq. \eqref{eq:Boltzmann}, for most standard species (including DM and baryons) it is possible to perform a Legendre decomposition and integrate analytically over the particle momentum $p$. Subsequently, one can evolve integrated quantities such as the density fluctuation $\delta$ and the divergence of fluid velocity $\theta$. 

In the presence of scattering processes between DM and massless neutrinos, the perturbation equations for the DM fluid are governed by the following equations:
\begin{align}
& \dot{\delta}_{\mathrm{DM}}=-\theta_{\mathrm{DM}}+3 \dot{\phi} 
\label{eq:DM_delta}
\\
& \dot{\theta}_{\mathrm{DM}}=k^2 \psi-\mathcal{H} \theta_{\mathrm{DM}}-\frac{4}{3}\frac{\rho_{\nu}}{\rho_{\mathrm{DM}}}\, \dot{\mu}\left(\theta_{\mathrm{DM}}-\theta_{\nu}\right),
\label{eq:DM_theta}
\end{align}
where $\mathcal{H}=\dot a /a$, while $\delta_{\mathrm{DM}}$ and $\theta_{\mathrm{DM}}$ are the DM density fluctuation and the divergence of fluid velocity, respectively. Here, $\dot{\mu}=a\,n_{\mathrm{DM}}\,\sigma_{\nu\mathrm{DM}}$ represents the $\nu$DM interaction rate defined in terms of the DM number density $n_{\mathrm{DM}}=\rho_{\mathrm{DM}} / m_{\mathrm{DM}}$ (i.e., the ratio between the DM energy density and the mass of DM particles) and the $\nu$DM cross-section $\sigma_{\nu\mathrm{DM}}$. Notice that, depending on the specific portal for $\nu$DM interactions, $\sigma_{\nu\mathrm{DM}}$ may have a temperature dependence. The two most common cases studied in the literature are $\sigma_{\nu\mathrm{DM}}\sim \mathrm{const}.$ and $\sigma_{\nu\mathrm{DM}}\sim T^2$. In this work, we will focus exclusively on the former case, deriving all results by assuming no temperature evolution in $\sigma_{\nu\mathrm{DM}}$.

For massless neutrinos, the modified Boltzmann hierarchy (following the same notation of Ref.~\cite{Ma:1995ey}, also for the definition of the harmonic expansion coefficients $F_{\nu,l}$ in the following equations) is given by:

\begin{align}
&\dot{\delta}_{\nu}  =-\frac{4}{3} \theta_{\nu}+4 \dot{\phi}\,, \\
&\dot{\theta}_{\nu}  =k^2\left(\frac{\delta_{\nu}}{4}-\sigma_{\nu}\right)+k^2 \psi-\dot{\mu}\left(\theta_{\nu}-\theta_{\mathrm{DM}}\right)\,, \\
& \dot{F}_{\nu, l}  =\frac{k}{2 l+1}\left[l F_{\nu, (l-1)}-(l+1) F_{\nu, (l+1)}\right]-\alpha_l \, \dot{\mu} \, F_{\nu, l}\,,\label{eq:generic_ell} 
%
\end{align}
with $\sigma_\nu=F_{\nu,2}/2$ the shear stress and Eq.~\eqref{eq:generic_ell} being valid for $l\geq 2$. The Boltzmann equation has been transformed into an infinite hierarchy of moment equations that must be truncated at some finite order $l_M$. We adopt the commonly used truncation scheme~\cite{Ma:1995ey},
\begin{equation}
F_{\nu,\left(l_{M}+1\right)} \approx \frac{\left(2 l_{M}+1\right)}{k \tau} F_{\nu, l_{M}}-F_{\nu,\left(l_{M}-1\right)}\,,
\label{eq:truncation_scheme}
\end{equation}
based on approximating ${F}_{\nu, l}(k\tau)\propto j_l(k\tau)$ and extrapolating the behavior to $l = (l_M + 1)$ using a recurrence relation for spherical Bessel functions $j_l(k\tau)$. Note that, as in the case of DM perturbations, $\nu$DM interactions alter the equations for $\theta_{\nu}$. In this case, they also affect the equations for $F_{\nu, l}$ with $l\geq2$. The modifications involve a set of numerical coefficients $\alpha_{l}=\mathcal O(1)$, whose exact values are determined by the dependence of the matrix element for the scattering process on the cosine of the angle
between the incoming and the outgoing neutrino.\footnote{In practice, following Ref.~\cite{Escudero:2015yka}, one can set $\alpha_2=9/5$ and $\alpha_{\ell}=1$ for $\ell \geq 3$ or determine the angular coefficients of the interaction terms of the higher-order multipoles as done in Ref.~\cite{Stadler:2019dii}. As shown in this latter work, the differences between the two cases are very small.} 

\subsection{\boldmath{$\nu$}DM Perturbations for massive neutrinos}
\label{sec.theory.massive}
Even though the massless approximation appears to effectively capture the phenomenology induced by $\nu$DM interactions, it appears by now to be established that neutrinos should be regarded as massive particles. For instance, the tightest lower limit set by neutrino oscillation experiments on the total neutrino mass at 95\% confidence level (CL) reads $\sum m_{\nu}\gtrsim 0.06$ eV~\cite{deSalas:2020pgw,Esteban:2020cvm,Capozzi:2021fjo}. In this study, we aim to get rid of this theoretical approximation and consider the more realistic case where neutrinos are considered massive particles. Therefore, here we extend the formalism presented in the previous subsection to this latter scenario.

In order to include the effect of massive neutrinos, we strictly follow Ref.~\cite{Mosbech:2020ahp}. Assuming neutrinos have a small but nonzero mass implies that they are neither inherently ultra-relativistic (like photons) nor non-relativistic (like baryons or DM). This holds true in presence of $\nu$DM interactions, as well. Therefore, one must solve the full hierarchy in a momentum-dependent manner. Using again the conventions of Ref.~\cite{Ma:1995ey}, for the non-interacting case, equations read: 
\begin{align}
& \dot{\Psi}_0=-\frac{p k}{E_\nu(p)} \Psi_1-\dot{\phi} \frac{d \ln f^{(0)}(p)}{d \ln p} \,,
\label{eq:noninteracting_massive_nu_1}
\\
&\dot{\Psi}_1=\frac{1}{3} \frac{p k}{E_\nu(p)}\left(\Psi_0-2 \Psi_2\right)-\frac{E_\nu(p) k}{3 p} \psi \frac{d \ln f^{(0)}(p)}{d \ln p} \,,
\label{eq:noninteracting_massive_nu_2}
\\
&\dot{\Psi}_l=\frac{1}{2 l+1} \frac{p k}{E_\nu(p)}\left(l \Psi_{l-1}-(l+1) \Psi_{l+1}\right), \quad l \geq 2
\label{eq:noninteracting_massive_nu_3}
\end{align}

For massive neutrinos, scatter-like interaction with DM produces an additional collision term in the Euler equation as well as a damping term to the higher momenta of the Boltzmann hierarchy. As a result, the final equations in presence of $\nu$DM interactions can be expressed as: 

\begin{align}
&\left[\dot{\Psi}_0\right]_{\nu\rm{DM}} = \dot{\Psi}_0\,,\\
&\left[\dot{\Psi}_1\right]_{\nu\rm{DM}} = \dot{\Psi}_1 -C_{\nu\rm{DM}} \left(\frac{\theta_{\rm{DM}} E_\nu(p)}{3k f^{(0)}(p)} \frac{d f^{(0)}(p)}{d p} + \Psi_1\right)\,,\\
&\left[\dot{\Psi}_2\right]_{\nu\rm{DM}}  =\dot{\Psi}_2-\frac{9}{10} C_{\nu\rm{DM}} \Psi_2\,, \\
&\left[\dot{\Psi}_l\right]_{\nu\rm{DM}}  =\dot{\Psi}_l-C_{\nu\rm{DM}} \Psi_l, \quad l \geq 3\,,
\end{align}
where $\dot{\Psi}_l$ are the standard parts given by Eqs. \eqref{eq:noninteracting_massive_nu_1} - \eqref{eq:noninteracting_massive_nu_3}, 
and 
\begin{equation}
C_{\nu\rm{DM}} = a\, \sigma_{\nu\rm{DM}} \left(\frac{n_{\rm DM} \, p^2}{E_{\nu}^2(p)}\right) 
\end{equation}
is a moment-dependent interaction rate. In this case we truncate the Boltzmann hierarchy as follows,

\begin{equation}
\left[\dot{\Psi}_{l_M+1}\right]_{\nu\rm{DM}}=\frac{(2l_M+1)E_\nu(p)}{p k \tau}\left[\dot{\Psi}_{l_M}\right]_{\nu\rm{DM}}-\left[\dot{\Psi}_{l_M-1}\right]_{\nu\rm{DM}}\,.
\end{equation}
The corresponding Boltzmann equations for DM can be obtained by integrating over momenta and be expressed in the following form:
\begin{align}
& \dot{\delta}_{\mathrm{DM}}=-\theta_{\mathrm{DM}}+3 \dot{\phi}\,, 
\\
& \dot{\theta}_{\mathrm{DM}}=k^2 \psi-\mathcal{H} \theta_{\mathrm{DM}}- \left(1+w_{\nu}\right)\frac{\rho_{\nu}}{\rho_{\mathrm{DM}}}\, \dot{\mu}\left(\theta_{\mathrm{DM}}-\theta_{\nu}\right)\,.
\label{eq:DM_theta_massive}
\end{align}
When compared to the massless case, we can see that the only difference arises from Eq. \eqref{eq:DM_theta_massive}, where in the momentum conservation factor $w_{\nu}=P_{\nu} / \rho_\nu$ is no longer always equal to $1/3$. Additionally, it is worth stressing that, unlike the massless case, here $\dot{\mu}$ does not have a simple analytical expression while keeping the same physical meaning. For additional details and discussions, we refer to Ref.~\cite{Mosbech:2020ahp}.

\subsection{\boldmath{$\nu$}DM Phenomenology}
By incorporating equations for perturbations into the most widely used Boltzmann solver codes, one can precisely integrate them and calculate the effects of $\nu$DM interactions on cosmological observables. As a result, numerous constraints on $\nu$DM interactions have emerged from a wealth of cosmological and astrophysical observations. Typically, these constraints are formulated in terms of a single dimensionless parameter
\begin{equation}
u_{\nu\mathrm{DM}} \doteq \left[\frac{\sigma_{\nu\mathrm{DM}}}{\sigma_{\mathrm Th}}\right] \left[\frac{m_{\mathrm DM}}{100\, \mathrm{GeV}}\right]^{-1}
\label{eq:u}
\end{equation}
that essentially quantifies the strength of the interaction normalizing the $\nu$DM cross section to the Thomson scattering rate and the mass of DM particles in units of 100 GeV. Since the value of $u_{\nu\mathrm{DM}}$ determines the collisional damping scale, as we argue in this section, depending on its value, the imprints left by $\nu$DM interactions can be different and manifest themselves at different scales.

One of the most significant $\nu$DM effects (which is also the primary focus of this work) is observed in the CMB angular power spectra of temperature and polarization anisotropies. To gain a better physical understanding of the phenomenology induced by $\nu$DM interactions, we can naively note that we expect them to influence the behavior of DM and neutrinos during both the radiation-dominated and matter-dominated phases. In fact, the shape of the CMB angular spectra is sensitive to the gravitational forces experienced by the coupled photon-baryon fluid before decoupling. These gravitational forces, in turn, are determined by free-streaming neutrinos and DM. In the standard scenario without interactions, when breaking down the solution to the cosmological perturbation system into slow and fast evolving modes, we observe that photon-baryon and neutrino perturbations are characterized as fast modes, while DM perturbations fall into the category of slow modes. Consequently, the photon-baryon fluid exhibits significant gravitational interactions only with free streaming neutrinos. More precisely, during the era of radiation domination and shortly after crossing the Hubble horizon, photon-baryon perturbations undergo a gravitational amplification which is gradually reduced due to free streaming neutrinos that develop anisotropic stress and cluster less effectively compared to a relativistic perfect fluid.\footnote{Note that modes that cross the Hubble radius during the matter-dominated epoch do not undergo this effect since the gravitational potential remains constant, while DM perturbations grow proportionally with the expansion of the universe $\delta_{\rm{DM}}\propto a$.} In the presence of interactions, the situation becomes quite different since DM experiences damped oscillations similar to neutrinos rather than undergoing slow gravitational clustering. Consequently, DM perturbations also contribute to the fast modes. In addition, neutrinos, being coupled to DM particles, do not free-stream anymore and their anisotropic stress is reduced, causing them to behave more similarly to a relativistic perfect fluid. As a result, the gravitational boost of photon-baryon perturbations increases. Similarly, if DM is still efficiently coupled to neutrinos when perturbations cross the Hubble radius during the matter-dominated era before recombination, being  gravitationally coupled to the photon-baryon fluid, it contributes to the fast mode solution. This leads to another gravitational boosting effect. All these differences in the dynamics of perturbations may lead to various effects on the CMB spectra, which have been extensively examined in the literature. Without aiming to provide a comprehensive treatment, it is worth mentioning, for example, that the enhanced gravitational boost felt by photon-baryon perturbations in the radiation-dominated epoch can potentially amplify all acoustic peaks except the first one while the second boost experienced during the matter-dominated Universe can amplify the first acoustic peak, as well. Other effects involve the fact that if DM is still efficiently coupled to neutrinos at the time of photon decoupling, metric fluctuations can get strongly suppressed, slightly enhancing even peaks and suppressing odd peaks. Additionally, $\nu$DM-fluid typically has a lower sound speed compared to the baryon-photon fluid, and this can produce a shift in the acoustic peaks towards slightly larger $\ell$. For more detailed discussions see Refs.~\cite{Wilkinson:2014ksa,Wilkinson:2013kia}.

Another widely studied signature arising from $\nu$DM interactions, is the significant impact on the matter power spectrum~\cite{Mangano:2006mp,Serra:2009uu,Wilkinson:2014ksa,Stadler:2019dii,Mosbech:2020ahp}. Interactions lead to an effective $\nu$DM fluid with non-zero pressure, and this pressure induces diffusion-damped oscillations in the matter power spectrum, analogous to the acoustic oscillations in the baryon-photon fluid. Therefore, the most remarkable effect on the matter power spectrum is a suppression of power on small scales. The characteristic scale of damped oscillations ($k_{\rm d}$) depends on the strength of the interactions: larger couplings will correspond to later epochs of neutrino-DM decoupling and a regime of damped oscillation on larger scales. Conversely, for smaller interaction strengths, the suppression becomes relevant on much smaller scales (higher wavenumber $k$). This has been documented in several studies involving both massless and massive neutrinos and was clearly pointed out in Ref.~\cite{Mangano:2006mp}, where -- see their Eq. (22) -- a proportionality $k_{\rm d} \propto (1/u_{\nu\rm{DM}})^{1/2}$ was derived between the wavenumber associated with the diffusion-damped oscillations and the intensity of interactions in the case of temperature-independent cross-section.

In general, this suppression of power on small scales can also influence the CMB angular spectra. Always without claiming to be exhaustive, we note that in the high multipole regime $\ell \sim \mathcal O(10^3)$, the spectrum of temperature anisotropies becomes directly proportional to the lensing power spectrum (see, for instance, Eq.(4.16) in Ref.~\cite{Lewis:2006fu}), which, in turn, depends on the distribution of matter in the Universe.\footnote{Hence, we expect the inclusion of the reconstructed lensing power spectrum of the CMB (i.e., the four-point function) in data analysis to be of great significance.} Furthermore, the matter power spectrum influences the growth of cosmic structures over time, which has an impact on the distribution of galaxies and, consequently, on the integrated Sachs-Wolfe effect. This contributes to CMB anisotropies caused by the gravitational evolution of structures. For this reason, we expect that as the strength of the interaction decreases, the effects on the CMB spectra should primarily manifest at smaller scales, akin to what occurs in the matter power spectrum.\footnote{Given the direct proportionality between $k$ and $\ell$, it is reasonable to expect that for lower coupling values, these effects will intensify with increasing $\ell$, becoming noticeable at higher $\ell$ values.}  This was argued in Refs.~\cite{Brax:2023rrf,Brax:2023tvn}, where it was also shown that including small-scale CMB measurements can be crucial for revealing unique signatures from interactions that would be challenging to detect on larger angular scales.

Finally, we conclude this section with a last important remark: in this scattering-type interaction between neutrinos and DM (at least to first order), no direct differences are expected at the level of background expansion. This is very different from, e.g., models of DM-DE interactions which modify the background expansion as well.

\subsection{Cosmological Model}
\label{sec:Cosmology}
We extended the cosmological model to include $\nu$DM interactions and determine the constraints that can be obtained from different combinations of the latest CMB and large-scale structure data. Specifically, we consider two different cosmological parameterizations.
\begin{itemize}
\item \textit{Massless neutrinos:} We begin with the most commonly used parameterization in the study of these interactions, where neutrinos are treated as massless and ultra-relativistic particles in the early Universe. This approximation is widely employed because it simplifies the equations for perturbations while capturing the overall $\nu$DM phenomenology quite accurately. Specifically, we employ the equations discussed in \autoref{sec.theory.massless}. In our analysis, we account for the interaction between neutrinos and the entire fraction of DM energy density, setting the effective number of ultra-relativistic particles at recombination ($N_{\rm eff}$) to its reference value of $N_{\rm eff}=3.044$. As demonstrated in Refs.~\cite{Brax:2023rrf,Brax:2023tvn}, no significant differences are observed when relaxing this latter assumption. It is important to note that considering neutrinos as massless particles is clearly an approximation and that we will get rid of it in the work. However, it is instructive to start with this simple case since it allows us to test some \textit{very} mild hints in favor of interactions documented in the literature (which mostly involve this approximation, see Refs.~\cite{Brax:2023rrf,Brax:2023tvn}), thereby corroborating existing results with a Profile Likelihood analysis and extending them to different combinations of data. Hopefully, this will also facilitate a better understanding of the results obtained when considering the most realistic case where neutrinos are regarded as massive particles.

\item \textit{Massive neutrinos:} We then extend the model to the case where neutrinos can be considered massive particles adopting the formalism detailed in \autoref{sec.theory.massive}. Despite the common practice in the standard cosmological model to fix the total neutrino mass to $\sum m_{\nu}=0.06$ eV, in our study, we proceed with full generality and leave it as a free parameter to be constrained by data.\footnote{Fixing $\sum m_{\nu} = 0.06$ eV and comparing the CMB spectra for the massless and the massive case, we tested afterward that the differences are minimal, regardless of the value of $u_{\nu\rm{DM}}$. This is predictable since one expects to recover the massless regime in the small mass limit. However, the differences between the two cases can substantially grow for larger mass values. This is the reason why we decided to keep $\sum m_{\nu}$ free to vary.} Specifically, we assume that all families of neutrinos interact with DM under the same interaction strength, we set the effective number of relativistic degrees of freedom at recombination to $N_{\rm eff}=3.044$, and we consider interactions between neutrinos and the entire fraction of DM. 

\end{itemize}
In both cases, we compute the theoretical model and introduce the possibility of interactions between neutrinos and DM using a modified version of the Cosmic Linear Anisotropy Solving System code, \texttt{CLASS}~\cite{Blas:2011rf}.\footnote{A publicly available version of this modified \texttt{CLASS} can be found at \url{https://github.com/MarkMos/CLASS_nu-DM}, see also Refs.~\cite{Stadler:2019dii, Mosbech:2020ahp}.}

\section{Methodology and data}
\label{sec:method}

\begin{table}
	\begin{center}
		\renewcommand{\arraystretch}{1.5}
		\begin{tabular}{c@{\hspace{0. cm}}@{\hspace{0.3 cm}} c @{\hspace{0.3cm}} c }
			\hline
			\textbf{Parameter}                       & \textbf{Massless Neutrinos}  & \textbf{Massive Neutrinos}\\
			\hline\hline
			$\Omega_{\rm b} h^2$                     & $[0.005\,,\,0.1]$ &$[0.005\,,\,0.1]$\\
			$\omega_{\rm{dm}}\equiv\Omega_{\rm c}^{\nu\rm DM} h^2$         & $[0.005\,,\,0.1]$ &$[0.005\,,\,0.1]$\\
			$100\,\theta_{\rm {MC}}$                 & $[0.5\,,\,10]$ &$[0.5\,,\,10]$\\
			$\tau$                                   & $[0.01\,,\,0.8]$ & $[0.01\,,\,0.8]$\\
			$\log(10^{10}A_{\rm S})$                 & $[1.61\,,\,3.91]$ & $[1.61\,,\,3.91]$\\
			$n_{s}$                                  & $[0.8\,,\, 1.2]$ & $[0.8\,,\, 1.2]$\\
			$\log_{10}u_{\nu\rm{DM}}$                  & $[-8\,,\,-1]$ & $[-8\,,\,-1]$\\
            $\sum m_{\nu}$ [eV]                            & -- & $[0.06\,,\,10]$\\
			
			\hline\hline
		\end{tabular}
		\caption{List of the uniform parameter priors. In some of our analyses we also employ a Gaussian prior for $\tau$ with a width much smaller than the uniform prior reported in this table, which depends on the concrete CMB data set under study. We provide the mean values and standard deviations of these Gaussian priors in \autoref{sec:Data}.}
		\label{tab.Priors}
	\end{center}
	
\end{table}

\subsection{Markov Chain Monte Carlo}
\label{sec:MCMC}
In order to perform the MCMC analysis, we make use of the publicly available sampler \texttt{COBAYA}~\cite{Torrado:2020xyz}. The code explores the posterior distributions of a given parameter space using the MCMC sampler developed for \texttt{CosmoMC}~\cite{Lewis:2002ah} and tailored for parameter spaces with speed hierarchy, implementing the "fast dragging" procedure detailed in Ref.~\cite{Neal:2005}. Our baseline sampling considers the six $\Lambda$CDM parameters, namely the baryon $\omega_{\rm b}\doteq \Omega_{\rm b}h^2$ and cold dark matter $\omega_{\rm dm}\doteq\Omega_{\rm c}^{\nu\rm DM} h^2$ energy densities, the angular size of the horizon at the last scattering surface $\theta_{\rm{MC}}$, the optical depth $\tau$, the amplitude of primordial scalar perturbation $\log(10^{10}A_{\rm s})$, and the scalar spectral index $n_s$. In addition, we consider the logarithm of the coupling parameter $\log_{10}u_{\nu \rm{DM}}$ -- where $u_{\nu \rm{DM}}$ is defined in Eq.~\eqref{eq:u} -- and perform a logarithmic sample to cover several orders of magnitude. Finally, when regarding neutrinos as massive particles, we include the total neutrino mass $\sum m_{\nu}$ as a free parameter as well, imposing the lower bound $\sum m_\nu>0.06$ that is found in neutrino oscillation experiments. This is important to get tighter and more realistic constraints on this parameter~\cite{Loureiro:2018pdz}. The prior distributions for all the sampled parameters involved in our analysis are chosen to be uniform along the range of variation provided in \autoref{tab.Priors}, with the only exception of the optical depth at reionization $\tau$ for which the prior distribution is chosen according to the CMB datasets discussed in \autoref{sec:Data}. Apart from the sampling cosmological parameters listed above and the nuisance parameters used to model the theory and the CMB experiment
systematics, we also obtain constraints on some important derived parameters, as $H_0$, the root-mean-square of mass fluctuations at the scale $R_8=8h^{-1}$Mpc, $\sigma_8$, and $S_8=\sigma_8(\Omega_m/0.3)^{0.5}$. The convergence of the chains obtained with this procedure is tested using the Gelman-Rubin criterion~\cite{Gelman:1992zz}. The threshold for chain convergence may need to vary from case to case to ensure sufficiently low levels of noise in the Profile Likelihoods. For most models, a good threshold for convergence has been found to be $R-1 \lesssim 0.02$ , while for some cases, a better convergence has been needed (all the way up to $R-1 \lesssim 0.001$).

\subsection{Marginalization and Profile Likelihood Analyses}
\label{sec:PL}

The most straightforward way of extracting constraints for a particular subset of  parameters of a model 
from the corresponding Monte Carlo Markov Chains is by applying the so-called marginalization procedure, which basically consists in integrating the original posterior distribution over those parameters we are not interested in. If the parameters of our model are contained in the vector $\theta=\{\theta_1,\theta_2\}$, with $\theta_1$ and $\theta_2$ two non-intersecting subsets of parameters, the marginalized distribution for $\theta_1$ simply reads, 

\begin{equation}
\mathcal{P}(\theta_1) = \int \mathcal{P}(\theta_1,\theta_2)\,d\theta_2\,.
\end{equation}
In practice, this integral can be trivially computed from the chains just by binning $\theta_1$ and counting the number of points in the chain that fall in each bin. This can be done pretty fast. The resulting histogram is an approximation of the marginalized distribution $\mathcal{P}(\theta_1)$. Better convergence of the chains leads to more accurate results, of course. Although this method is very efficient, it can introduce biases due to volume (a.k.a marginalization) effects when the original posterior distribution has important non-Gaussian features, as it is clearly illustrated in Ref.~\cite{Gomez-Valent:2022hkb}. 

Profile Likelihoods stand as a good alternative to the marginalized distributions. The former can be regarded as being complementary to the latter because they carry different information. The main advantage of the Profile Likelihood is that it is not subject to volume effects, since it does not involve any integration. The Profile Likelihood associated to $\theta_1$ is defined as follows, 

\begin{equation}\label{eq:PD}
\tilde{\mathcal{P}}(\theta_1)=\max\limits_{{\theta_2}}\mathcal{P}(\theta_1,\theta_2)\,,
\end{equation}   
up to a normalization factor. Despite the simplicity of this expression, its computation can be quite time-consuming due to the fact that it requires to bin $\theta_1$ and perform the maximization of the original posterior distribution over $\theta_2$ in each of these bins.  This is why the vast majority of the Profile Likelihood analyses in cosmology have focused on only one or two parameters, see e.g.~\cite{Herold:2021ksg,Reeves:2022aoi,Holm:2022kkd,Herold:2022iib,Holm:2023laa,Cruz:2023cxy,McDonough:2023qcu,Efstathiou:2023fbn}. However, Ref.~\cite{Gomez-Valent:2022hkb} showed that it is also possible to compute the Profile Likelihood directly from the chains and with enough precision to assess the impact of volume effects in the usual marginalization process. This is done much faster, even if in some occasions larger chains are needed to decrease the noise. This method allowed to obtain the full set of one-dimensional marginalized and Profile Likelihood constraints for the main and derived parameters of the $\Lambda$CDM and several models beyond it~\cite{Gomez-Valent:2022hkb}, and has been already applied in other works, as in Ref.~\cite{Galloni:2022mok}.

The use of Profile Likelihoods is important to detect biases in the interpretation of the output of standard Monte Carlo analyses, which are in general based on marginalization. For instance, Refs.~\cite{Herold:2021ksg,Gomez-Valent:2022hkb,Herold:2022iib,Efstathiou:2023fbn} showed that in the context of the ultra-light axion-like early dark energy models, volume effects might play a non-negligible role, producing important shifts in several parameters that are pivotal in the discussion of the $H_0$ tension. Quantifying these biases is always useful (even in $\Lambda$CDM~\cite{Planck:2013nga,Gomez-Valent:2022hkb}), but it becomes particularly relevant when discussing new physics because they can influence our conclusions.

In this paper we apply the methodology of Ref.~\cite{Gomez-Valent:2022hkb} to compute the one-dimensional Profile Likelihoods for the models with $\nu$DM interactions explained in \autoref{sec:theory}, using the datasets described in the next section. We will compare these results with those obtained with the corresponding marginalized distributions, which we generate making use of the \texttt{Python} package \texttt{GetDist} \cite{Lewis:2019xzd}. This will allow us to test the robustness of the results reported in Refs.~\cite{Brax:2023rrf,Brax:2023tvn}.

\subsection{Cosmological Data}
\label{sec:Data}
Our reference datasets in the study of $\nu$DM interactions are the following:
\begin{itemize}
    \item The Planck 2018 temperature and polarization (TT TE EE) likelihood, which also includes low multipole data ($\ell < 30$)~\cite{Planck:2019nip,Planck:2018vyg,Planck:2018nkj} and the Planck 2018 lensing likelihood~\cite{Planck:2018lbu}, constructed from measurements of the power spectrum of the lensing potential. We refer to this dataset as \textit{\textbf{P18}}.
    
    \item Atacama Cosmology Telescope DR4 likelihood, combined with a Gaussian prior on $\tau = 0.065 \pm 0.015$, as done in~\cite{ACT:2020gnv}. We refer to this dataset as \textit{\textbf{ACT-DR4}}.
    
    \item  The gravitational lensing mass map covering 9400 deg$^2$ reconstructed from CMB measurements made by the Atacama Cosmology Telescope from 2017 to 2021~\cite{ACT:2023kun,ACT:2023dou}. In our analysis we include only the conservative range of lensing multipoles $40 < \ell < 763$.   We refer to this dataset as \textit{\textbf{ACT-DR6}}.

    \item The South Pole Telescope temperature and polarization (TT TE EE) likelihood~\cite{SPT-3G:2021eoc,SPT-3G:2022hvq} combined with a Gaussian prior on $\tau = 0.065 \pm 0.015$.  We refer to this dataset as \textit{\textbf{SPT}}.
    
    \item Baryon acoustic oscillations (BAO) measurements extracted from data from the 6dFGS~\cite{Beutler:2011hx}, SDSS MGS~\cite{Ross:2014qpa} and BOSS DR12~\cite{BOSS:2016wmc} surveys. We refer to this dataset as \textit{\textbf{BAO}}.
\end{itemize}

Notice that we will often consider combinations of different CMB experiments. Most notably, we will combine Planck data at large angular scales with ACT or SPT data at small scales. The reason for doing so is that both ACT and SPT lack data around the first acoustic peaks. In the absence of such data, a large degeneracy among cosmological parameters typically arises.\footnote{This is the case, for instance, with parameters like $n_s$ and $\Omega_b\, h^2$ whose constraints in the absence of large-scale data do not always agree with the ones derived considering comprehensive multipole coverage~\cite{ACT:2020gnv,DiValentino:2022rdg,Giare:2022rvg,Giare:2023wzl,DiValentino:2022oon,Calderon:2023obf,Giare:2023xoc}.} Additionally, precise measurements of the first peaks are crucial in determining the total neutrino mass. As it is well known, when light neutrinos switch from a relativistic to a non-relativistic regime, they alter the gravitational potentials and leave characteristic signatures in the CMB angular power spectra through the integrated Sachs-Wolfe effect. Such effects are typically remarkable on the multipole range missed by ground-based telescopes.\footnote{For this reason, cosmological bounds on the total neutrino mass derived from ground-based telescope data are typically much more relaxed, exceeding the eV value. See, e.g., Refs.~\cite{DiValentino:2021imh,DiValentino:2023fei}.} On the other hand, as pointed out previously in this work, small-scale measurements provided by ground-based telescopes can be crucial to probe small $\nu$DM coupling values. In general, when combining different CMB experiments, we will ensure that we consider Planck data only in the multipole range not covered by the other experiment. For instance, when considering the high-$\ell$ ACT multipoles in combination with Planck data, we will use Planck data only in the multipole range between $2 \lesssim \ell \lesssim 650$. This is necessary to avoid including the region where the two experiments overlap, which would result in double counting of the same sky patch if a covariance matrix is not considered~\cite{ACT:2020gnv}. Similar considerations apply when combining SPT and Planck data.

\begin{table*}
\begin{center}
\renewcommand{\arraystretch}{1.1}
\resizebox{0.7\textwidth}{!}{ \begin{tabular}{l@{\hspace{0. cm}}@{\hspace{1.5 cm}} c @{\hspace{1.5cm}} c }

\toprule
\nq{\textbf{Dataset}}  & \nq{\textbf{Massless Neutrinos} \\ ${\log_{10} u_{\nu\text{DM}}}$}  & \nq{\textbf{Massive Neutrinos} \\ ${\log_{10} u_{\nu\text{DM}}}$}\\
\hline\hline
&&\\
\textbf{P18+BAO} & \nq{Mar: $<-4.27$\\ PL: $<-4.34$} & \nq{Mar: $-4.11^{+0.73}_{-0.93}$\\ PL: $-5.00^{+0.90}_{-1.80}$} \\
\vspace{0.05cm}\\
\hline
&&\\
\textbf{ACT-DR4+BAO} & \nq{Mar: $-4.12^{+0.49}_{-0.90}$\\ PL: $-4.17^{+0.58}_{-0.87}$} & \nq{Mar: $-4.05^{+0.94}_{-1.21}$\\ PL: $-3.90^{+0.65}_{-1.25}$} \\
&&\\
\textbf{ACT-DR4+DR6+BAO}  & \nq{Mar: $-4.35^{+0.52}_{-0.79}$\\ PL: $-4.37^{+0.48}_{-0.80}$} & \nq{Mar: $-4.12^{+0.68}_{-1.32}$\\ PL: $-4.00^{+0.59}_{-0.91}$} \\
&&\\
\textbf{ACT-DR4+P18+BAO} & \nq{Mar: $-4.64^{+0.60}_{-0.67}$\\ PL: $-4.60^{+0.46}_{-0.58}$} & \nq{Mar: $-4.19^{+0.39}_{-0.45}$\\ PL:$-3.96^{+0.44}_{-0.66}$} \\
&&\\
\hline
&&\\
\textbf{SPT+BAO} & \nq{Mar: $<-3.56$\\ PL: $<-3.51$} & \nq{Mar: $<-3.15$\\ PL: $-4.6^{+1.1}_{-1.7}$} \\
&&\\
\textbf{SPT+P18+BAO} & \nq{Mar: $<-3.90$\\ PL: $-4.58^{+0.46}_{-2.04}$} & \nq{Mar: $-5.5\pm 1.2$\\ PL: $-5.7\pm 1.2$} \\
\bottomrule
\end{tabular}}
\caption{Results for $\log_{10}(u_{\nu{\rm DM}})$ in the scenarios of massless and massive neutrinos derived from both marginalized posteriors and Profile Likelihoods. The datasets used are combinations of different CMB and BAO data described in \autoref{sec:method}. The central values correspond to the peaks of their respective one-dimensional distributions, and all confidence intervals are reported at $68$\% CL. Conversely, the upper bounds are consistently reported at $95$\% CL. We employ the conservative lower bound $\log_{10}(u_{\nu{\rm DM}})=-8$ in the computation of the distribution normalization factors.}
\label{tab.Results}
\end{center}
	
\end{table*}

\section{Results for Massless Neutrinos}\label{sec:results_massless}

\label{Results.massless}
In this section, we present the results obtained under the approximation of massless neutrinos. It is important to emphasize once again (as the saying goes, "repetita iuvant") that although many of the results we will mention in this section have already been addressed in the literature (see, e.g., recent Refs.~\cite{Brax:2023rrf,Brax:2023tvn}), this analysis aims to further validate these findings using different statistical methodologies such as the Profile Likelihood analysis (PL hereafter) and to extend the discussion to other experiments and data that have not been explored in relation to these models (such as the ACT-DR6 lensing data and the SPT data). Given the large amount of data and experiments to consider, in order to promote better organization and clarity, we will proceed by dividing this section into different subsections, each dedicated to discussing in detail the results obtained from a specific CMB experiment and eventually its combinations with others. Finally, we will dedicate the last subsection to point out some concluding remarks, condense the main findings of our analysis, and provide a takeaway message. Note also that, in order to keep the discussion of the results clear and concise, we will only present results and plots related to the parameters of interest for this model, see \autoref{tab.Results} and Figs. \ref{fig:Massive_vs_Massless} and \ref{fig:comaprison_logu}. Additional Tables containing the results for all cosmological parameters, their correlations in the form of contour plots, as well as a detailed parameter-by-parameter comparison between the PL and the marginalized posterior distributions are provided in the supplementary material. More concretely, they are presented in \autoref{tab:table_masless} and Figs. \ref{fig:nomass_P18}-\ref{fig:SPT+BAO_full}. We encourage the reader to consult the supplementary materials whether they wish to obtain a more complete overview of the results.

\subsection{Planck}
\begin{figure*}[t!]  
\includegraphics[scale=0.65]{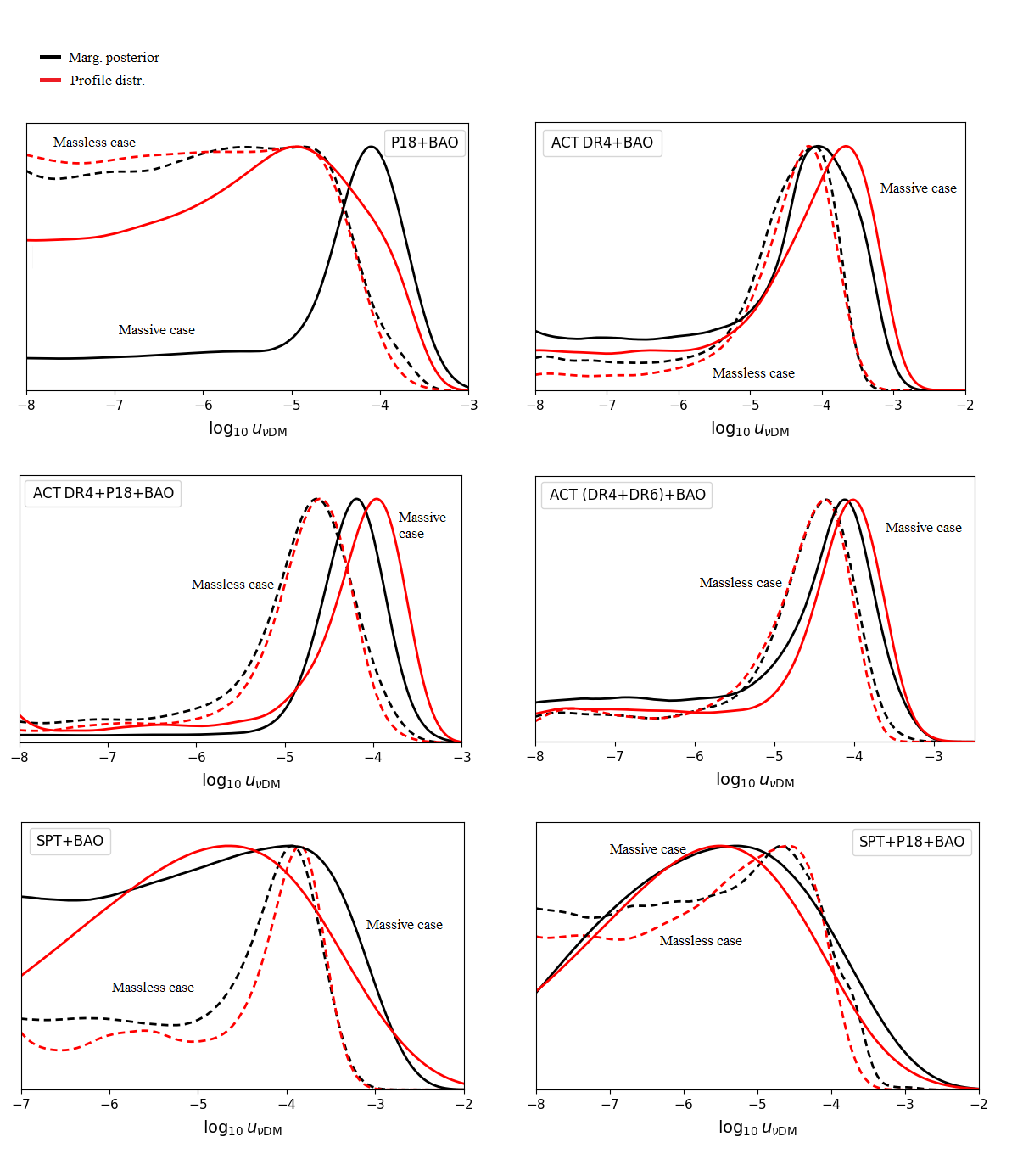}    \caption{Comparison of the one-dimensional marginalized  and profile distributions of $\log_{10}u_{\nu{\rm DM}}$ (in black and red, respectively) considering massless neutrinos (with dashed lines) and massive neutrinos (with solid lines), for each of the datasets considered in this paper.}\label{fig:Massive_vs_Massless}
\end{figure*}

Let us begin our analysis with the Planck 2018 data. We recall that we take into account both temperature and polarization measurements, as well as the lensing spectrum reconstruction. In addition, we consider BAO measurements, referring to the final combination of data as P18+BAO. In this case, no clear preference for $\nu$DM interaction of any kind is found. This is consistently confirmed both through a marginalization analysis and a PL analysis. Using both methodologies, we get a 95\% CL upper limit on the parameter quantifying the interaction strength:
\begin{align}
\text{Marg: } & \log_{10}u_{\nu\rm{DM}} < -4.27 \\
\text{PL: }   &  \log_{10}u_{\nu\rm{DM}} < -4.34
\end{align}
The one-dimensional marginalized and profile distributions for this parameter are depicted in the top-left plot of \autoref{fig:Massive_vs_Massless}. From the figure, it is evident that there are no strong preferences for interactions, and when the value of $\nu$DM coupling becomes very small, the probability distributions become flat. The PL analysis makes it clear that this happens because models with small interactions do not exhibit significant differences in the $\chi^2$ of the fit and are essentially indistinguishable from each other, offering equally valid descriptions of the data. The negligible impact of volume effects is corroborated by \autoref{fig:P18+BAO_full} and \autoref{tab:table_masless} for the other cosmological parameters, as well. The differences between the marginalized and PL distributions (and, hence, between the constraints derived from them) are minimal.

\subsection{Atacama Cosmology Telescope}
A relatively recent development in the study of $\nu$DM interactions emerged when it was noticed that, for weak interaction strengths, effects left by interactions can be orders of magnitude more pronounced in the damping tail than at larger angular scales. As a result, the inclusion of small-scale CMB data can significantly enhance our ability to constrain these models. Interestingly, analyzing temperature and polarization measurements from the Atacama Cosmology Telescope (which covers angular scales up to approximately $\ell\sim4200$, significantly smaller than those explored by the Planck satellite), a modest preference for interactions has been noted~\cite{Brax:2023rrf,Brax:2023tvn}.

First and foremost, we underscore that this preference is further validated in light of the present reanalysis. When considering the ACT-DR4 data in combination with the BAO data, the results we obtain at 68\% CL read:
\begin{align}
\text{Marg: } & \log_{10}u_{\nu\rm{DM}} = -4.12^{+0.49}_{-0.90} \\
\text{PL: }   &  \log_{10}u_{\nu\rm{DM}} = -4.17^{+0.58}_{-0.87} 
\end{align}
Of significant importance is the fact that the PL analysis confirms the preference for $\nu$DM interactions, unequivocally demonstrating that it arises by an actual decrease of the $\chi^2$ value of the fit and showing a perfect agreement with the results obtained from the marginalized distribution. This improvement essentially happens in the region of large multipoles ($\ell>2000$) of the ACT-DR4 TTTEEE likelihood, see Appendix \ref{sec:AppendixB}.

Secondly, we take another step forward in the study of these intriguing signals by expanding our analysis to include the recent ACT-DR6 lensing likelihood~\cite{ACT:2023kun, ACT:2023dou}. As already mentioned, the different behavior of neutrino free-streaming in the presence of $\nu$DM interactions, as well as the broader phenomenology described in \autoref{sec:theory}, can result in a reduction of power in the matter spectrum at small scales, thereby leaving imprints on the lensing potential. In addition, the spectrum of temperature anisotropies at high-$\ell$ becomes proportional to the lensing spectrum. All these reasons lead us to anticipate that ongoing efforts in the reconstruction of the DM distribution through the lensing spectrum could potentially improve our constraints on interactions. Considering ACT-DR4 in conjunction with ACT-DR6 and BAO data, we get:
\begin{align}
\text{Marg: } & \log_{10}u_{\nu\rm{DM}} = -4.35^{+0.52}_{-0.79} \\
\text{PL: }   &  \log_{10}u_{\nu\rm{DM}} = -4.37^{+0.48}_{-0.80}
\end{align}
at 68\% CL. Remarkably, such a hint of interaction remains stable when including lensing data and considering the PL distribution.

Finally, we check whether combining temperature and polarization data from ACT with Planck data at larger angular scales preserves this preference intact. As mentioned earlier, if DM is still efficiently coupled to neutrinos when perturbations cross the Hubble radius during the matter-dominated era, an amplification of the first peak of the spectrum of temperature anisotropies may be induced. Since ACT temperature measurements lack data around the first acoustic peaks (they probe a range of multipoles $\ell \gtrsim 600$ in TT), we may not be able to accurately constrain this possibility, potentially missing crucial information. By considering the combination of ACT-DR4+Planck+BAO (where Planck is considered only in the multipole range where it does not overlap with ACT), we eventually obtain at 68\% CL:
\begin{align}
\text{Marg: } & \log_{10}u_{\nu\rm{DM}} = -4.64^{+0.60}_{-0.67} \\
\text{PL: }   &  \log_{10}u_{\nu\rm{DM}} = -4.60^{+0.46}_{-0.58}
\end{align}
We notice that when combining large and small-scale data, not only does the preference for interactions persist, but it becomes even slightly more pronounced due to a reduction in uncertainties. Once again, the PL analysis confirms that this preference arises from a global improvement in the fit to small scales without worsening the fit to larger-scale CMB data.

In light of these findings, we can confidentially conclude that a mild preference for interaction $\log_{10} u_{\nu\rm{DM}}\sim -4.5$ is found in ACT data. 
Indeed, in all the cases considered, over 68\% of the total probability distribution is contained within the peaked range of $-5 \lesssim \log_{10} u_{\nu\text{DM}} \lesssim -3$, while the remaining probability lies in the long left tail $\log_{10}u_{\nu\rm{DM}} \lesssim -5$. This results in a non-negligible detection of $\nu$DM at a confidence level exceeding 68\%, but falling short of the 95\% confidence level. Alternatively, this corresponds to a statistical significance ranging between one and two standard deviations.
As clear from \autoref{fig:ACT+lensing+BAO_full}, the fact that this hint disappears when considering the 95\% CL results is due to our choice to maintain a highly conservative approach and perform logarithmic sampling with a very wide prior for the interaction strength, covering several orders of magnitude down to $\log_{10} u_{\nu\rm{DM}}\sim -8$. Clearly, for such exceedingly modest values of coupling, the effects of interactions become negligible at any scale, resulting in flat tails of the distributions, both for the marginalized and profile distributions. This latter makes it clear that the minimum $\chi^2$ stops changing as this parameter changes. The dependence of the prior on the results obtained in this study certainly warrants careful consideration, and we encourage the interested reader to consult Appendix \ref{sec:AppendixA} for further details.

\subsection{South Pole Telescope}
We conclude our analysis of $\nu$DM interactions in the limit of massless neutrinos by extending our study to the South Pole Telescope data. It is worth noting that, up to this point, SPT temperature and polarization measurements have never been examined in the context of this interacting model. Nonetheless, despite larger uncertainties compared to other CMB experiments, SPT still provides sufficiently precise data to (at least try to) independently test the results obtained by Planck and ACT.

As usual, we start by combining SPT and BAO. When doing so, we get only an upper bound on the interaction strength which at 95\% CL reads
\begin{align}
\text{Marg: } & \log_{10}u_{\nu\rm{DM}} < -3.56, \\
\text{PL: }   &  \log_{10}u_{\nu\rm{DM}} < -3.51.
\end{align}
Clearly, when it comes to SPT, the increasing uncertainties make it challenging to derive definitive conclusions. In spite of this, it is worth looking at the PL and the marginalized probability distribution functions given in \autoref{fig:SPT+BAO_full}. Choosing to see the bright side, we can speculate about the presence of a modest peak at $\log_{10}u_{\nu\rm{DM}}\sim -4$, which seems to point more in the direction indicated by ACT than Planck. The same mild preference is confirmed by the PL distribution: the fact that this peak does not disappear reassures us that it is given by a genuine improvement in the $\chi^2$ and not by volume effects. However, the fraction of probability contained below the peak is $\sim 60\%$ and therefore it does not reach the level of one standard deviation. One must also consider that we are adopting a very conservative large prior range for the parameter quantifying interactions. Smaller prior ranges would lead to larger levels of significance. Once more, we encourage referring to Appendix~\ref{sec:AppendixA} for further details.

That being said, any SPT preference for interactions appears to be reduced compared to ACT. This result was somewhat expected as SPT probes multipoles up to $\ell \sim 3000$ (comparable to Planck) without delving into the small angular scales reached by ACT. 

\begin{figure}[t!]  
\includegraphics[width=\columnwidth]{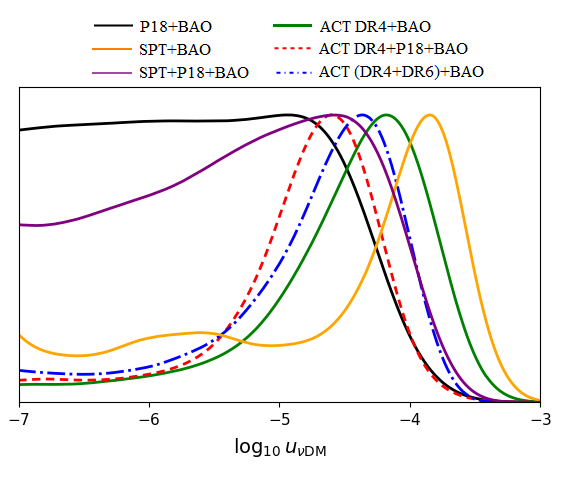}    \caption{Comparison of the profile distributions of $\log_{10}u_{\nu{\rm DM}}$ obtained with the various datasets described in \autoref{sec:Data} and considering massless neutrinos. See \autoref{sec:results_massless} for details.}\label{fig:comaprison_logu}
\end{figure}

Another aspect to take into account is that SPT shares similar limitations with ACT, particularly regarding the lack of data around the first acoustic peaks in the spectrum of temperature and polarization anisotropies. As already explained, such a deficiency can produce additional loss of information when constraining this model. For this reason, we perform the same test detailed in the previous subsection and combine SPT with Planck in order to cover the missing multipole range while avoiding overlap between the two experiments. Considering SPT+P18+BAO, we obtain:
\begin{align}
\text{Marg: } & \log_{10}u_{\nu\rm{DM}} <-3.90\quad (95\%\,{\rm CL})\,, \\
\text{PL: }   &  \log_{10}u_{\nu\rm{DM}} = -4.58^{+0.46}_{-2.04}\quad (68\%\,{\rm CL})\,.
\end{align}
We observe that combining SPT with Planck the preference mentioned for SPT+BAO translates into an actual indication at 68\% CL in the PL that aligns perfectly with the results obtained from ACT (with or without Planck). Nevertheless, it is important to notice that in this case we obtain much larger uncertainties than when combining ACT and Planck and the peak in the marginalized and PL distributions is also much broader. This can be understood looking at \autoref{fig:comaprison_logu}. The most preferred values of $\log_{10}u_{\nu{\rm DM}}$ according to SPT+BAO fall a bit away from the plateau of the PL distribution obtained with P18+BAO. This explains why the signal weakens when we combine SPT and Planck. However, it is still present and in full agreement with the results obtained with ACT.

Overall, when considering these results individually, (even paying attention to these minor peaks) one may conclude that they do not convincingly support the hypothesis of interactions in the SPT data or that their statistical significance is certainly not sufficient to draw any conclusions. Anyway, if we step back and look at the bigger picture, it appears interesting that models with $u_{\nu\text{DM}}\sim 10^{-5}$ to $10^{-4}$ could potentially lead to modest improvements in fitting data from two out of three independent experiments as unequivocally confirmed by the PL analysis.

\subsection{Concluding Remarks}
In this section, in light of different statistical methods and additional experiments, we reviewed and reassessed some recently emerged hints suggesting a very mild preference for $\nu$DM interactions. Specifically, we confirmed that the Planck satellite data do not exhibit any distinct preference towards interactions, while ACT CMB measurements at smaller angular scales (where the impact of weak couplings are expected to be more pronounced) hint at a slight preference, ranging between one and two standard deviations. Notably, this preference persists including the recent ACT-DR6 lensing measurements, as well as combining ACT and Planck together. Given the limited level of statistical significance, such indications could potentially result from statistical fluctuations. However, the PL analysis confirms that our results are linked to a genuine improvement in the overall fit when $u_{\nu\text{DM}}\sim 10^{-5} - 10^{-4}$. Although to a much lesser extent and with an increased level of ambiguity, one may speculate about the presence of similar indications also in the SPT data. However, in this latter case, the larger experimental uncertainties make it challenging to derive any reliable conclusion.
\section{Results for Massive Neutrinos}
\label{sec:results_massive}

We make no secret that the results derived in the previous section, although intriguing, appear to be quite scattered. Clearly, an important element of uncertainty derives from assuming neutrinos as massless particles. With the goal of getting rid of this approximation and minimizing the number of assumptions, in this section, we replicate the same analysis considering neutrinos as massive particles and leaving their total mass a free parameter to be constrained by data. We maintain the same structure presented in the previous section, dividing the narration experiment by experiment and concluding with some general remarks. The main results are presented again in \autoref{tab.Results} and \autoref{fig:Massive_vs_Massless}. Additional details and results can be found in the supplementary material, in \autoref{tab:table_massive} and Figs. \ref{fig:P18_massive_contours}-\ref{fig:SPT+BAO_massive}.

\subsection{Planck}

Much like the massless scenario, our initial step involves combining Planck and BAO data. When focusing on this particular combination (which usually sets a precision and reliability standard among the highest attainable), we get what might be considered the most interesting result of this paper:
\begin{align}
\text{Marg: } & \log_{10}u_{\nu\rm{DM}} = -4.11^{+0.73}_{-0.93} \\
\text{PL: }   &  \log_{10}u_{\nu\rm{DM}} =-5.00^{+0.90}_{-1.80} 
\end{align}
Surprisingly, for massive neutrinos, we find a marginalized posterior distribution function for $\log_{10}u_{\nu{\rm DM}}$ that clearly peaks at coupling values consistent with those favored by small-scale CMB experiments. However this feature, which for the same dataset was absent in the massless limit, is not fully validated by the PL analysis, see the top-left plot in \autoref{fig:Massive_vs_Massless}. In particular, the profile distribution confirms an overall improvement in the $\chi^2$ statistics for interacting models, but the distribution is much wider and it does not appear to be peaked at the same values obtained by the marginalized distribution. Therefore, while the results obtained by the two methods are not in tension, a clear confirmation of this preference is missing. As usual, another element of uncertainty is that this indication nearly vanishes at the 68\% CL, due to the same prior-dependent effects mentioned in the case of massless neutrinos. Specifically, as the coupling becomes small, the tail of the posterior distribution becomes flat because the effects of interactions become subdominant at all scales. We once again encourage the reader to refer to Appendix \ref{sec:AppendixA} for further details.

For the sake of comprehensiveness, we also explore how neutrino mass bounds change in the presence of interactions. The limits we get at $95\%$ CL are\footnote{We do not explicitly show the lower bound $\sum m_\nu>0.06$ eV in any of the results reported in this paper. However, as shown in \autoref{tab.Priors}, this lower bound  holds in all the analyses performed in \autoref{sec:results_massive}. It is motivated by the results obtained in neutrino oscillation experiments~\cite{deSalas:2020pgw,Esteban:2020cvm,Capozzi:2021fjo}.}:
\begin{align}
\text{Marg: } & \sum m_{\nu} < 0.15 \, \rm{eV}\,,\\
\text{PL: }   & \sum m_{\nu} < 0.15 \, \rm{eV}\,.
\end{align}
By comparing the results presented in Tables \ref{tab:table_masless}-\ref{tab:table_massive} we can see that our constraints are not substantially different from those obtained in the previous section assuming massless neutrinos. Current cosmic data have sufficient power to constrain $\log_{10}u_{\nu{\rm DM}}$ and $\sum m_\nu$  independently so that leaving the total neutrino mass as a free parameter does not produce strong degeneracy with the interaction strength. Our results resonate well with those reported in~\cite{Mosbech:2020ahp}.

\subsection{Atacama Cosmology Telescope}
Turning now to the study of small-scale CMB measurements, an important result is that the same preference for interactions observed in ACT under the assumption of massless neutrinos persists when considering neutrinos as massive particles. Indeed, considering ACT-DR4+BAO for the interaction strength we obtain
\begin{align}
\text{Marg: } & \log_{10}u_{\nu\rm{DM}} = -4.05^{+0.94}_{-1.21}\,, \\
\text{PL: }   &  \log_{10}u_{\nu\rm{DM}} = -3.90^{+0.65}_{-1.25}\,,
\end{align}
at 68\% CL while the limits on the total neutrino mass read
\begin{align}
\text{Marg: } & \sum m_{\nu} < 0.39 \, \,\rm{eV}\,,\\
\text{PL: }   & \sum m_{\nu} < 0.36 \, \rm{eV}\,,
\end{align}
at 95\% CL. We observe a persistent indication of interactions, although now the uncertainties on the interaction parameter are much larger compared to the massless case. As for $\sum m_{\nu}$, in principle a nonzero neutrino mass can impact on the damping tail, potentially resulting in correlations with the effects of interaction. However, our constraints on $\sum m_{\nu}$ are predominantly driven by BAO data that effectively eliminates any degeneracy. This can be easily understood by noting that the results obtained on the total neutrino mass only from small-scale CMB data typically exceed the eV value~\cite{DiValentino:2023fei,DiValentino:2021imh}. 

The next step is to incorporate the latest ACT-DR6 lensing data into our analysis. When considering the dataset ACT-DR4+DR6+BAO, we obtain
\begin{align}
\text{Marg: } & \log_{10}u_{\nu\rm{DM}} = -4.12^{+0.68}_{-1.32}\,,  \\
\text{PL: }   &  \log_{10}u_{\nu\rm{DM}} = -4.00^{+0.59}_{-0.91}\,,
\end{align}
at 68\% CL and
\begin{align}
\text{Marg: } & \sum m_{\nu} < 0.24 \, \rm{eV}\,,\\
\text{PL: }   & \sum m_{\nu} < 0.24 \, \rm{eV}\,,
\end{align}
at 95\% CL. As already demonstrated for massless neutrinos, lensing data can significantly increase the precision obtained when constraining cosmological parameters, including $\nu$DM interactions. This holds true also for the massive case. For this dataset, the preference for interactions remains unchanged as highlighted by both the marginalized and Profile distributions in \autoref{fig:Massive_vs_Massless}. Furthermore, we observe an overall reduction of error bars in cosmological parameters, including the upper limit obtained on neutrino mass. This is consistent with our intuition that precise measurements of lensing effects (here captured by the recent ACT-DR6 lensing likelihood) can play a significant role in advancing our understanding of thermal relics and their properties~\cite{Giare:2023aix}.  

Finally, we conclude by considering the combination ACT-DR4+P18+BAO. In this case the results for $u_{\nu\rm{DM}}$ are 
\begin{align}
\text{Marg: } & \log_{10}u_{\nu\rm{DM}} = -4.19^{+0.39}_{-0.45}\,, \\
\text{PL: }   &  \log_{10}u_{\nu\rm{DM}} = -3.96^{+0.44}_{-0.66}\,,
\end{align}
at 68\% CL while for and $\sum m_{\nu}$ we get
\begin{align}
\text{Marg: } & \sum m_{\nu} < 0.18 \, \rm{eV}\,,\\
\text{PL: }   & \sum m_{\nu} < 0.17 \, \rm{eV}\,,
\end{align}
at 95\% CL. For this combination of data the preference for interactions becomes more pronounced compared to any other case analyzed in this study. As we can observe in \autoref{fig:Massive_vs_Massless}, the central part of the probability distribution is very narrow, while the tails decay rapidly, though they always remain flat for very low values of $\log_{10}u_{\nu\rm{DM}}$. This means that within the 95\% CL, a null (or very low) coupling between neutrinos and DM is consistent with the data, although interaction values on the order of $u_{\nu\rm{DM}}\sim 10^{-5} - 10^{-4}$ seem to be favored by both the marginalized and profile distributions. Regarding the mass of neutrinos, the constraint we obtain is the tightest one for the combination of data involving ACT. This is not surprising and is primarily due to the inclusion of data at large angular scales and specifically around the first acoustic peaks. Indeed it is well known that the most prominent impact of neutrino masses on CMB anisotropies occurs through the early integrated Sachs-Wolfe effect. Neutrinos may become non-relativistic around recombination, influencing gravitational potentials and leaving a distinctive signature that reaches its maximum around the first peak in the spectrum of temperature anisotropies, see e.g.~\cite{Lesgourgues:2012uu}.

We end this subsection by commenting on the observed systematic shift of the peaks of all the distributions of $\log_{10}u_{\nu{\rm DM}}$ when ACT and/or Planck are employed in the analyses, compared to those obtained considering massless neutrinos. This can be easily grasped in \autoref{fig:Massive_vs_Massless} and \autoref{tab.Results}. These shifts are not strong enough to introduce a significant tension between the results obtained in the two scenarios. They remain actually compatible at $\lesssim 1\sigma$ CL. 

\subsection{South Pole Telescope}

We conclude by analyzing the SPT data. When we consider SPT+BAO we obtain at $95\%$ CL
\begin{align}
\text{Marg: } & \log_{10}u_{\nu\rm{DM}} <-3.15\,, \\
\text{PL: }   &  \log_{10}u_{\nu\rm{DM}} =-4.6^{+1.1}_{-1.7}\,,
\end{align}
and 
\begin{align}
\text{Marg: } & \sum m_{\nu} <0.41 \, \rm{eV}\,,\\
\text{PL: }   & \sum m_{\nu} < 0.37 \, \rm{eV}\,,
\end{align}
for the interaction strength and the neutrino mass, respectively. As argued in the massless case, SPT currently exhibits experimental uncertainties greater than other CMB experiments, making it challenging to detect any preference for interactions. This limitation becomes even more severe here when neutrinos are considered massive, and their mass is treated as an additional free parameter in the cosmological model. However, taking these results alongside the one-dimensional marginalized and profile posterior distributions in \autoref{fig:Massive_vs_Massless} at face value, there is no inclination towards interaction. Regarding the constraint on the total neutrino mass, once again, most of the constraining power arises from BAO measurements.

Adding Planck data at large angular scales and considering the combination SPT+P18+BAO, our results read:
\begin{align}
\text{Marg: } & \log_{10}u_{\nu\rm{DM}} =-5.5\pm 1.2\,,\\
\text{PL: }   &  \log_{10}u_{\nu\rm{DM}} = -5.7\pm 1.2\,,
\end{align}
at 68\% CL for the interaction strength, and
\begin{align}
\text{Marg: } & \sum m_{\nu} < 0.19 \, \rm{eV}\,,\\
\text{PL: }   & \sum m_{\nu} < 0.17 \, \rm{eV}\,,
\end{align}
at 95\% CL for the total neutrino mass. The same considerations discussed in detail in the manuscript apply equally to the constraints on $u_{\nu\rm{DM}}$ and $\sum m_{\nu}$. In the case of the former, the SPT data does not provide compelling evidence in support of the interaction hypothesis. This is crystal-clear from the bottom-right plot of Fig. \ref{fig:Massive_vs_Massless}. Regarding the total neutrino mass, the incorporation of data covering large angular scales significantly improves our ability to constrain it.

\subsection{Concluding Remarks}
In this section, we have improved the analysis of $\nu$DM interactions by considering neutrinos as massive particles, with their mass regarded as a free parameter in the cosmological model. The most intriguing result we have obtained is a slight preference for interactions when considering only Planck and BAO data which is in good agreement with results obtained from ACT. However, this preference (which for the same dataset was absent approximating neutrinos as massless particles), is only partially confirmed by a PL analysis. In turn, ACT data continue to exhibit the same mild preference for  $u_{\nu\rm{DM}}\sim10^{-5} - 10^{-4}$, which remains consistent and more pronounced when we include gravitational lensing data and combine ACT with Planck. In particular, this preference becomes very clear for ACT-DR4+P18+BAO, although without substantially exceeding the level of 2 standard deviations. On the other hand, SPT data (either on its own or in conjunction with Planck) do not provide compelling evidence supporting interactions. However, due to the larger uncertainties, interactions are not disfavored either in the range preferred by the other CMB probes. Overall, we can conclude that \textit{all} the hints for interactions found in the massless limit persist when we relax this assumption and new consistent hints emerge (e.g., Planck+BAO). Consequently, while the situation remains undoubtedly open, it is not unreasonable to conclude that, based on all currently available CMB experiments and BAO data, a scatter-like interaction between neutrinos and DM with a strength $u_{\nu\rm{DM}}\sim 10^{-5}-10^{-4}$ appears to offer a modest improvement in fitting data.

\section{Conclusions}
\label{sec:conclusions}

A crucial and widely debated aspect in our current understanding of the Universe, is whether DM interacts with the other Standard Model particles, beyond gravitational interactions. Various possible interaction channels have been tested and studied in both cosmology and particle physics, including interactions with photons, baryons, dark radiation, and neutrinos.

Despite the entirety of these studies not providing convincing evidence supporting DM interactions, very recently, a few scattered results have been reported, hinting at a slight preference towards DM scattering with neutrinos when considering some independent collections of cosmological data. More precisely, in Ref.~\cite{Hooper:2021rjc} a preference for $\nu$DM interactions at the level of $\sim 3\sigma$ has been found by analyzing Lyman-$\alpha$ data. Subsequently, in Ref.~\cite{Brax:2023rrf, Brax:2023tvn} an independent yet consistent preference (at a statistical significance ranging between one and two standard deviations) emerged by analyzing CMB measurements of temperature and polarization anisotropies at small angular scales where the effects of tiny interactions have been argued to be significantly larger than on larger scales. 

Given the limited level of statistical significance, such indications could potentially result from statistical fluctuations and/or systematic effects in the data. For sure, at present, they do not appear substantial enough to assert compelling evidence for $\nu$DM interactions. In this regard, it is  also important to consider that analyses of other astrophysical and cosmological observations, such as the galaxy luminosity function, impose stringent constraints on interactions ($u_{\nu\text{DM}}\lesssim 10^{-6}$)~\cite{Mosbech:2022nkk} that are only marginally compatible with the aforementioned results. Additionally, incorporating interactions beyond the Standard Model encounters the typical challenges compounded by constraints from particle physics, introducing additional layers of complexity, see, e.g., the discussion in Refs.~\cite{Brax:2023rrf,Brax:2023tvn}.  That being said, we believe it is equally imperative not to overlook these intriguing signals that deserve at least further investigation and rigorous cross-checking to be fully understood. This is particularly true for indications pointing toward interactions in the CMB. Unlike late-time constraints, they are not strongly influenced by the nonlinear evolution of structures (which may introduce dependencies on specific models) but are based on a solid understanding of the underlying physics, where the dynamics of perturbations can be accurately described within a linear regime.

In this paper, we have undertaken a comprehensive re-evaluation of all hints emerged from CMB observations to conclusively determine their strength and credibility. In particular, we have analyzed all available CMB experiments in combination with Baryon Acoustic Oscillations measurements, focusing on the most recent temperature, polarization and gravitational lensing data released by Planck, ACT, and SPT. In order to test the results previously reported in the literature, for all these experiments, our analysis begins by considering neutrinos as massless particles (an assumption upon which the vast majority of available results are based). Subsequently, we have extended the analysis to the more general case in which neutrinos are treated as massive particles whose mass is left free to be determined by data. To ensure rigorous cross-check of the results and keep control over possible effects related to the vast volume of the parameter space (i.e., to avoid false detection or missing detection resulting from correlations among parameters), each combination of data has been analyzed both in terms of the usual marginalized probability distributions and by means of a Profile Likelihood methodology.

While the statistical significance of our results remains somewhat limited due to the current data sensitivity, our findings show a remarkable resilience in the above mentioned hints for interactions. In fact, their resilience becomes evident through cross-validation in independent experiments as well as in the consistency observed when comparing marginalized and profile distributions. In what follows we summarize (experiment by experiment) what we consider the most relevant and novel results of this manuscript:
\begin{itemize}

    \item \textbf{Planck:} The analysis of Planck temperature, polarization, and lensing data combined with Baryon Acoustic Oscillations measurements, does not exhibit a clear preference for $\nu$DM interactions. Considering neutrinos as massless particles, both the marginal distribution and the Profile Likelihood analysis show a flat posterior probability distribution function for values of the interaction strength $ u_{\nu\rm{DM}} \lesssim 10^{-4}$. From this, we can confirm that below this threshold value, Planck data are unable to distinguish between models with or without interactions. In the more realistic scenario where neutrinos are regarded as massive particles, the marginalized probability distribution gives a slight indication ($\sim 1\sigma$) in favor of interactions, but this indication is not fully confirmed by the PL analysis. However, the PL analysis does confirm that models with $ u \sim 10^{-5}-10^{-4}$ show a modest reduction in the $\chi^2$ value of the fit. Consequently, although there is not clear preference for $\nu$DM interactions, our reanalysis conclusively demonstrates that such scenarios are not in conflict with Planck data.
    
    \item \textbf{ACT:} Our analysis confirms the preference for $\nu$DM interactions that has recently emerged when including ACT data at small angular scales in the analysis~\cite{Brax:2023rrf,Brax:2023tvn}. In particular, we confirm that this preference persists at a level ranging between 1 and 2 standard deviations in all combinations of data involving ACT measurements at small scales. We observe a consistent preference for $ u_{\nu\rm{DM}} \sim 10^{-5}-10^{-4}$, whether considering only ACT-DR4 data related to temperature and polarization measurements, including the ACT-DR6 likelihood for the lensing spectrum, or considering ACT together with Planck data at larger angular scales. In all these cases, the preference persists both when neutrinos are considered massless particles and when they are considered massive particles. In addition, this preference is always supported by the Profile Likelihood analysis, which unequivocally demonstrates an overall improvement in the fit. It is important to note that in the more realistic case where neutrinos are considered to have mass, this preference becomes even more evident despite the broader parameter space.

    \item \textbf{SPT:} Since in Planck we do not find clear evidence for interactions, while in ACT we observe a moderate preference for interactions, a good method to discriminate between the two hypotheses is certainly to compare the results with a third independent experiment. To do so, we extend the analysis to the CMB measurements of the temperature and polarization anisotropies provided by SPT. Clearly, this experiment probes intermediate scales between ACT and Planck and currently exhibits error bars that are larger than both these experiments. Consequently, this makes it even more challenging to interpret the results. That being said, assuming neutrinos as massless particles, from SPT+BAO we observe that both the marginalized probability distribution and the profile distribution seem to indicate a slight preference for interactions. However, due to the large uncertainties and the conservative prior employed for the interaction strength, this preference does not reach the 68\% CL. The peak obtained with SPT+BAO is slightly separated from the plateau found with P18+BAO. This explains why if we consider large-scale data from Planck together with SPT+BAO, the preference weakens, giving wider peaks. Nevertheless, these results are fully compatible with ACT. 
Considering massive neutrinos (and thus introducing an additional parameter to constrain) leads to a further reduction in constraining power, which further complicates the interpretation of the results. 
  
    Overall, when considering SPT data individually, one may be tempted to conclude that they do not convincingly support the hypothesis of interactions and that their statistical significance is certainly not sufficient to draw reliable conclusions. While this is of course true, it is also true that models with $u_{\nu\text{DM}}\sim 10^{-5}$ to $10^{-4}$ are fully consistent with SPT and so that they are not disfavoured either.
\end{itemize}

In conclusion, there is a reasonable basis to hypothesise that an interaction resembling elastic scattering between neutrinos and DM, characterized by a strength of $u_{\nu\rm{DM}}\sim 10^{-5}-10^{-4}$, could potentially be supported by current CMB experiments, contributing to a modest improvement in the fit of data. Despite the undeniable uncertainty surrounding this hypothesis (which is, in turn, linked to the current limited experimental sensitivity), the consistency of the results observed across independent observations along with their resilience made evident by the Profile Likelihood analysis, provide intriguing hints that can be definitively tested in light of future experiments.

\begin{acknowledgments}
\noindent  AGV is supported by a fellowship from ``la Caixa'' Foundation (ID 100010434) with code LCF/BQ/PI23/11970027. He also receives support from grant  PID2022-136224NB-C21 funded by MCIN/AEI/10.13039/501100011033. EDV is supported by a Royal Society Dorothy Hodgkin Research Fellowship. CvdB is supported (in part) by the Lancaster–Manchester–Sheffield Consortium for Fundamental Physics under STFC grant: ST/X000621/1. This article is based upon work from COST Action CA21136 Addressing observational tensions in cosmology with systematics and fundamental physics (CosmoVerse) supported by COST (European Cooperation in Science and Technology). We acknowledge IT Services at The University of Sheffield for the provision of services for High Performance Computing.
\end{acknowledgments}
\appendix
\section{Prior dependence of \boldmath{$\log_{10}u_{\nu {\rm DM}}$}}\label{sec:AppendixA}

\begin{table*}[htpb!]
\centering
\resizebox{\textwidth}{!}{\begin{tabular}{|c ||c | c || c | c || c | c|| c | c|| c | c|}
 \multicolumn{1}{c}{} & \multicolumn{2}{c}{\textbf{P18+BAO}} & \multicolumn{2}{c}{\textbf{ACT DR4+P18+BAO}} &\multicolumn{2}{c}{\textbf{ACT (DR4+DR6)+BAO}} &\multicolumn{2}{c}{\textbf{SPT+BAO}} &\multicolumn{2}{c}{\textbf{SPT+P18+BAO}}
\\\hline
{\small $\log_{10}u_{\nu {\rm DM}}>$} & {\small Marg.}  & {\small PL} & {\small Marg.}  & {\small PL} & {\small Marg.}  & {\small PL} & {\small Marg.}  & {\small PL}  & {\small Marg.}  & {\small PL}
\\\hline
$-8$ & $<-4.27$ & $<-4.34$  & $-4.64^{+0.60}_{-0.67}$ & $-4.60^{+0.46}_{-0.58}$ & $-4.37^{+0.52}_{-0.77}$ & $-4.35^{+0.52}_{-0.79}$ & $<-3.56$ & $<-3.51$ & $<-3.90$ & $-4.58^{+0.46}_{-2.04}$ \\\hline
$-7$ & $<-4.17$ & $<-4.27$  & $-4.64^{+0.54}_{-0.59}$ & $-4.60^{+0.43}_{-0.53}$ & $-4.35^{+0.47}_{-0.66}$ & $-4.37^{+0.43}_{-0.66}$ & $<-3.53$ & $-3.84^{+0.46}_{-0.95}$  & $-4.75^{+0.39}_{-1.49}$ & $-4.58^{+0.48}_{-1.44}$ \\\hline
$-6$ & $<-4.04$ & $<-4.18$  & $-4.64^{+0.46}_{-0.51}$ & $-4.60^{+0.39}_{-0.47}$  & $-4.35^{+0.42}_{-0.56}$ & $-4.37^{+0.39}_{-0.57}$ & $<-3.50$ & $-3.84^{+0.38}_{-0.56}$  & $-4.75^{+0.45}_{-1.02}$ & $-4.58^{+0.44}_{-0.91}$  \\\hline
\end{tabular}}
\caption{Comparison between the results for massless neutrinos obtained for the parameter $\log_{10}(u_{\nu{\rm DM}})$ when we use the lower bounds $\log_{10}(u_{\nu{\rm DM}})=-8,-7,-6$ in the computation of the distribution normalization factors. Two-sided constraints are reported at 68\% CL and one-sided constraints at 95\% CL.}\label{tab:table_appendix}
\end{table*}

\begin{figure*}[t!]  
\includegraphics[width=0.9\textwidth]{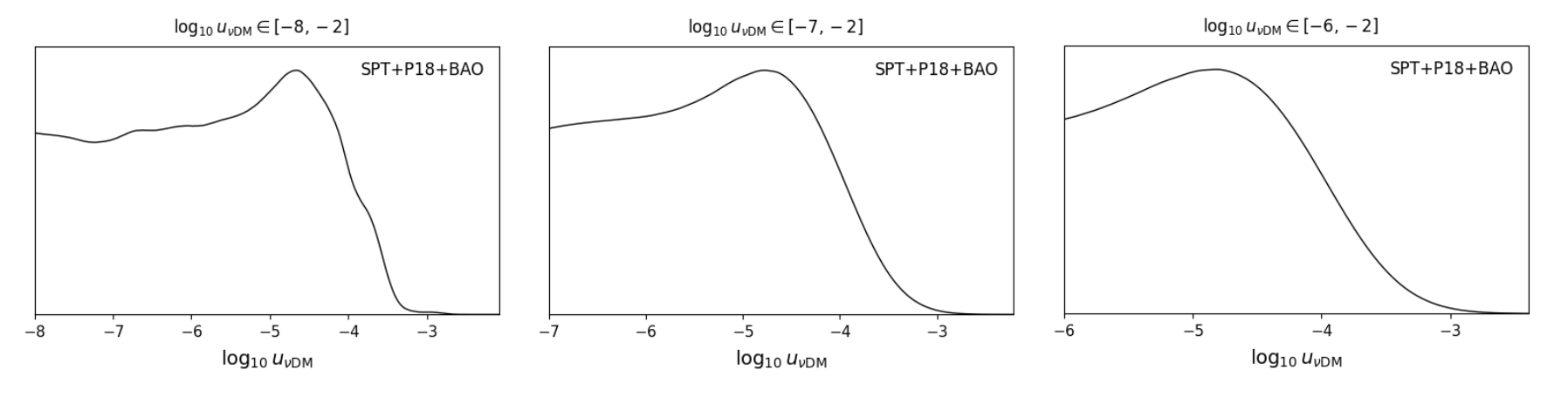}  
\caption{Comparison among the marginalized 1D posterior probability distribution functions for $\log_{10} u_{\nu \rm{DM}}$ obtained for SPT+P18+BAO in the massless neutrinos scenario considering three different priors on this parameter.}
\label{fig:prior}
\end{figure*}

We devote this brief appendix to discuss the impact of the lower bound of the flat prior employed for the interaction strength. As shown in \autoref{tab.Priors}, in all the analyses of this paper we have sampled this parameter in the range $\log_{10}u_{\nu {\rm DM}}\in [-8,-1]$. This is certainly a very conservative choice, since our datasets are essentially unable to distinguish interacting models with $u_{\nu {\rm DM}}\lesssim 10^{-6}$ from $\Lambda$CDM, which is formally recovered in the limit $u_{\nu {\rm DM}}\to 0$ (or $\log_{10}u_{\nu {\rm DM}}\to -\infty$). However, in the light of the constraining power of the data found in this study, the standard model is already retrieved when $u_{\nu {\rm DM}}\sim 10^{-6}$. This means that for the data under consideration all the points in the region $\log_{10}u_{\nu {\rm DM}}< -6$ correspond in practice to a $\Lambda$CDM model. Therefore, it is clear that we have been quite conservative in this study, giving a larger weight to the standard model due to the small lower bound employed in our prior of the coupling. The normalization factors of our distributions are larger than what they would have been if we had used a larger lower bound, of course, and this translates into slightly smaller evidences for the interaction between dark matter and neutrinos. For illustrative purposes, we show this explicitly in \autoref{tab:table_appendix}, where we compare the constraints we get on $\log_{10}u_{\nu {\rm DM}}$ when the lower bound of the prior is set to -8, -7 and -6, considering massless neutrinos. The constraints are tightened by $\sim 20-30\%$ in the analyses involving ACT data and by $\sim 50\%$ in the case of SPT+P18+BAO. Therefore, the choice of the prior, to some extent, can actually have a non-negligible impact on the constraints we get on the coupling. This is the reason why we are indeed choosing a very conservative prior.

To explicitly discuss why we believe the choice of the prior is very conservative, we consider a direct example and examine the SPT+P18+BAO dataset in \autoref{tab:table_appendix}. In this case, narrowing the prior for the coupling results in a significant difference in the outcomes, specifically a 68\% CL interval rather than an upper limit on $\log_{10} u_{\nu \rm{DM}}$.Therefore, for illustrative purposes, it is instructive to focus on this case and explain why we believe we have been very conservative in choosing our priors, reinforcing the robustness of our results.

The 1D posteriors (obtained from \texttt{GetDist}) for the coupling with three different priors are shown in \autoref{fig:prior}. As depicted in the figure, decreasing the prior range for the parameter has no effect other than increasing the fraction of total probability under the peak and reducing the fraction of probability in the left tail of the probability distribution. For this specific dataset, when considering a broad prior, the fraction of probability under the peak does not reach 68\%, and therefore, we can report an upper limit at 95\% CL. On the contrary, when we reduce the prior, the total probability fraction under the peak clearly increases, reaching 68\% CL. Consequently, it could be stated that the peak corresponds to a one standard deviation preference for non-vanishing interactions.

Therefore assuming a lower limit of $\min(\log_{10} u_{\nu \rm{DM}})=-8$ is a conservative choice as all indications are obtained within very broad priors for this parameter.

\section{The ACT preference for \boldmath{$\nu {\rm DM}$}}
\label{sec:AppendixB}

Based on the results presented in this paper and discussed in previous works on the same model~\cite{Brax:2023rrf,Brax:2023tvn}, we can conclude that the apparent preference for $\nu$DM interactions arises from the inclusion of ACT temperature and polarization data. In this regard, it is worth stressing one more time that, although probing a range of multipoles that partially overlaps with the Planck satellite measurements, ACT extends to much smaller angular scales (i.e., larger multipoles) than the latter. For instance, ACT has sensitivity in the spectrum of temperature anisotropies covering the multipole range $\ell \in [650, 4200]$, while Planck, for the same spectrum, probes scales $\ell \in [2, 2500]$. Since the effects of new physics in the cosmological model may manifest differently at various angular scales, small-scale CMB measurements may provide crucial insights when testing models beyond $\Lambda$CDM, serving as an independent test of the results derived from the Planck satellite data and extending them to scales not measured by the latter.

Interestingly, as emphasized in several recent studies, a few intriguing hints for new physics supporting the last statement have already surfaced from ACT data. To highlight two of the most interesting results, a preference at approximately 3 standard deviations in favor of a pre-recombination early dark energy has been pointed out in Ref.~\cite{Hill:2021yec} while in Ref.~\cite{Kreisch:2022zxp}, it was argued that, combining ACT data with large-scale information from WMAP, delayed onset of neutrino free streaming, possibly caused by significantly strong neutrino self-interaction, seems to be statistically favored at a level ranging from 2 to 3 standard deviations. In light of these results, we find it imperative and interesting to highlight how our conclusions fit into this context and eventually clarify the relation between the hint for $\nu$DM interactions discussed here and the other ACT indications for new physics beyond the standard cosmological model.

Firstly, we observe that, as noted in Ref.~\cite{Hill:2021yec}, the preference for EDE arises from the ACT TE and EE polarization measurements. On the other hand, in Ref.~\cite{Kreisch:2022zxp}, it has been shown that the potential preference for neutrino self-interaction can be attributed to the multipole range $700\lesssim \ell \lesssim 1000$ in the ACT E-mode polarization measurements. In both cases, the Planck satellite measures the multipole range (in temperature and polarization) from which the ACT preference for new physics emerges without, however, confirming these intriguing results. As argued in the same abstract of Ref.~\cite{Hill:2021yec}, tight constraints on early dark energy can be derived from Planck high-$\ell$ TT data alone, and no preference is found. Furthermore, when comparing the best-fit early dark energy spectra obtained for ACT and Planck, coherent differences emerge across a broad range of multipoles in TE and EE.

\begin{figure}[t!]  
\includegraphics[width=0.8\columnwidth]{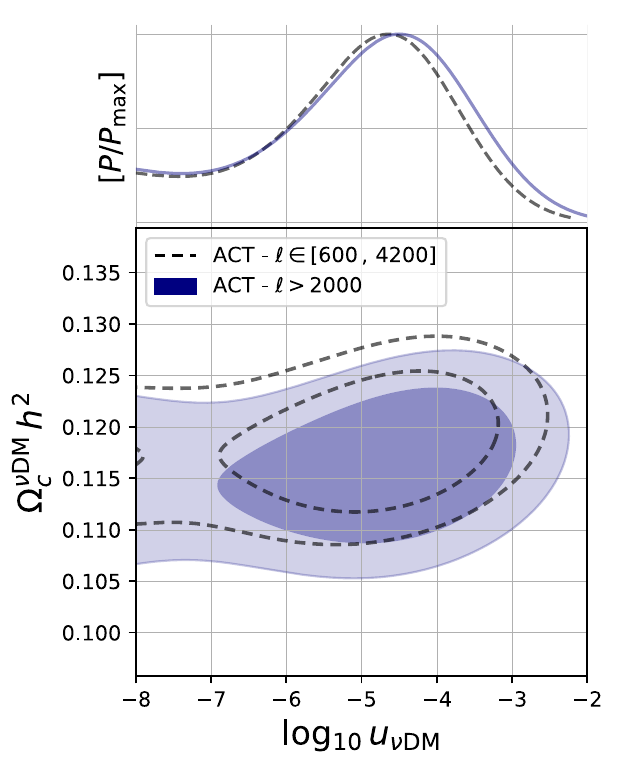}    
\caption{2D contours at 68\% and 95\% CL and 1D posterior probability distributions for the coupling parameter $\log_{10} u_{\nu \rm{DM}}$ and the total interacting cold dark matter energy density $\Omega^{\nu\rm{DM}}_{c}\,h^2$ obtained with the full ACT-DR4 TTTEEE likelihood (dashed grey contours) and only considering multiples $\ell\gtrsim 2000$ (full blue contours).}
\label{fig:ell}
\end{figure}

As argued in Refs.~\cite{Brax:2023rrf,Brax:2023tvn} (and as we aim to conclusively demonstrate in this appendix), this is not the case for $\nu$DM interactions. Indeed, as discussed in Ref.~\cite{Brax:2023rrf} and reiterated in this work, the effects of $\nu$DM coupling, at the order favored by ACT, can become significantly large at small scales while remaining quite modest at the scales measured by the Planck satellite. This has been explicitly shown in Figure 2 of Ref.~\cite{Brax:2023tvn}, where the angular power spectra of temperature anisotropies corresponding to the ACT and Planck best-fit values are compared. As clear from the figure, despite the preference arising from ACT, assuming a non-zero coupling does not lead to significant differences in the multipole range of Planck, contrary to what occurs, for example, with EDE. For this reason, the preference for $\nu$DM interactions was argued to arise from ACT temperature and polarization measurements at very small scales not measured by Planck.

However, to unequivocally prove the last claim, here we consider the likelihood of ACT-DR4, which includes measurements of temperature and polarization anisotropies, and derive constraints on $\nu$DM interactions under two scenarios: \textit{i)} analyzing the entire ACT-DR4 TTTEEE likelihood, and \textit{ii)} truncating the same likelihood into different bins, retaining only datapoints at $\ell>2000$. This second option in practice implies discarding all ACT temperature and polarization measurements at scales measured by the Planck satellite while keeping data points on small scales not probed by the latter. The results obtained for these two cases are shown in \autoref{fig:ell} where we present the 1D marginalized posterior for $\log_{10} u_{\nu\rm{DM}}$ and its 2D-correlation with $\Omega^{\nu\rm{DM}}_{c}$ (as extracted from our MCMC analysis using GetDist). 

Despite an expected widening of uncertainties, considering the entire ACT likelihood or only the part corresponding to $\ell>2000$,  does not lead to significant differences, and the same preference for $\nu$DM interactions is confirmed:

\begin{align}
\text{Full ACT-DR4: } & \log_{10}u_{\nu\rm{DM}} =  -5.03^{+1.30}_{-0.92}\,, \\
\text{ACT-DR4 } \ell> 2000:   &  \log_{10}u_{\nu\rm{DM}} =-4.9^{+1.5}_{-1.0}\,.
\end{align}
This unequivocally indicates that this preference largely arises from small scales $\ell\gtrsim 2000$.

\bibliographystyle{apsrev4-1}
\bibliography{Profile_nuDM}
\clearpage
\widetext

\begin{center}
\textbf{\Large \textsc{Supplementary Material}}
\end{center}

\section*{Massless Neutrinos}

\subsection*{Tables}

\begin{table*}[htbp!] 
\centering
\renewcommand{\arraystretch}{2}
\resizebox{\textwidth}{!}{
\begin{tabular}{|c ||c | c || c | c || c | c || c | c || c | c |}
 \multicolumn{1}{c}{} & \multicolumn{2}{c}{\textbf{P18+BAO}} & \multicolumn{2}{c}{\textbf{ACT DR4+P18+BAO}} &\multicolumn{2}{c}{\textbf{ACT (DR4+DR6)+BAO}} &\multicolumn{2}{c}{\textbf{SPT+BAO}} &\multicolumn{2}{c}{\textbf{SPT+P18+BAO}}
\\\hline
{\small Parameter} & {\small Marg.}  & {\small PL} & {\small Marg.}  & {\small PL} & {\small Marg.}  & {\small PL} & {\small Marg.}  & {\small PL} & {\small Marg.}  & {\small PL}
\\\hline
$100\omega_b$ & $2.239\pm 0.014$ & $2.247^{+0.012}_{-0.014}$ & $2.236\pm 0.012$ & $2.236^{+0.013}_{-0.012}$ & $2.158\pm 0.029$ & $2.167^{+0.025}_{-0.031}$ & $2.222\pm 0.031$ & $2.219^{+0.033}_{-0.028}$ & $2.234\pm 0.012$ & $2.238^{+0.010}_{-0.011}$ \\\hline
$10\omega_{\rm dm}$ & $1.196\pm-0.009$ & $1.194^{+0.009}_{-0.008}$ & $1.198\pm 0.010$ & $1.198\pm 0.009$ &  $1.199\pm 0.014$ 
 & $1.200\pm 0.013$ & $1.184\pm 0.015$ & $1.186^{+0.013}_{-0.015}$ & $1.196\pm 0.010$ & $1.198^{+0.007}_{-0.010}$ \\\hline
$H_0$ & $68.05\pm 0.42$ & $68.15^{+0.39}_{-0.41}$ & $68.00\pm 0.42$ & $67.97^{+0.43}_{-0.41}$ & $67.63\pm 0.55$ & $67.83^{+0.48}_{-0.57}$ & $68.06\pm 0.59$ & $68.28^{+0.38}_{-0.71}$ & $67.93\pm0.42$ & $67.95^{+0.42}_{-0.43}$ \\\hline
$\tau$ & $0.053\pm 0.007$& $0.054\pm 0.007$& $0.056\pm 0.007$ & $0.056\pm 0.006$ & $0.063^{+0.013}_{-0.011}$ & $0.069^{+0.010}_{-0.014}$   & $0.056\pm 0.014$  & $0.058^{+0.011}_{-0.013}$ & $0.054\pm 0.007$ & $0.054^{+0.005}_{-0.006}$ \\\hline
$n_s$ & $0.965\pm 0.004$ & $0.966\pm 0.004$ & $0.968\pm 0.004$ & $0.968^{+0.004}_{-0.003}$  & $0.997\pm 0.012$ & $0.998^{+0.012}_{-0.011}$ & $0.964\pm 0.016$ & $0.967^{+0.014}_{-0.015}$ & $0.965\pm 0.004$ & $0.966\pm 0.004$ \\\hline
$\ln(10^{10}A_s)$ & $3.040\pm 0.014$ & $3.043\pm 0.013$ & $3.053\pm 0.013$ & $3.055\pm 0.013$ & $3.054\pm 0.021$ & $3.061^{+0.020}_{-0.023}$ & $3.043\pm 0.030$ & $3.045^{+0.031}_{-0.025}$ & $3.041\pm 0.013$ & $3.045^{+0.011}_{-0.013}$ \\\hline
$\log_{10}u_{\nu {\rm DM}}$ & $<-4.27$ & $<-4.34$  & $-4.64^{+0.60}_{-0.67}$ & $-4.60^{+0.46}_{-0.58}$ & $-4.35^{+0.52}_{-0.79}$ & $-4.37^{+0.48}_{-0.80}$ & $<-3.56$ & $<-3.51$ & $<-3.90$ & $-4.58^{+0.46}_{-2.04}$ \\\hline
$\sigma_8$ & $0.820^{+0.007}_{-0.008}$  & $0.822^{+0.007}_{-0.008}$  & $0.821^{+0.010}_{-0.011}$ & $0.820^{+0.009}_{-0.011}$ & $0.830^{+0.013}_{-0.012}$ & $0.831^{+0.011}_{-0.016}$ & $0.811^{+0.018}_{-0.029}$ & $0.805^{+0.018}_{-0.035}$ & $0.820^{+0.009}_{-0.010}$ & $0.823^{+0.008}_{-0.011}$ \\\hline
$S_8$ & $0.827^{+0.013}_{-0.012}$ & $0.829^{+0.011}_{-0.014}$ & $0.830^{+0.015}_{-0.014}$ & $0.831^{+0.013}_{-0.015}$  & $0.841^{+0.016}_{-0.014}$  & $0.843^{+0.013}_{-0.015}$ & $0.812^{+0.025}_{-0.031}$  & $0.798^{+0.031}_{-0.035}$ & $0.829^{+0.014}_{-0.016}$ & $0.838^{+0.010}_{-0.015}$ \\\hline
\end{tabular}
}
\caption{Fitting results obtained with the various CMB datasets described in \autoref{sec:method} and employing both, the marginalized posteriors and Profile Likelihoods under the approximation of massless neutrinos. The central values refer to the location of the peaks of the corresponding one-dimensional distributions and all the two-sided constraints are quoted at $68\%$ CL, while one-sided constraints are provided at 95 \% CL. For the parameter $\log_{10}(u_{\nu{\rm DM}})$ we employ the conservative lower bound $\log_{10}(u_{\nu{\rm DM}})=-8$ in the computation of the distribution normalization factors.}
\label{tab:table_masless}
\end{table*}

\clearpage 
\subsection*{Correlations among cosmological parameters}

\begin{figure*}[htbp!]  
\includegraphics[width=\textwidth]{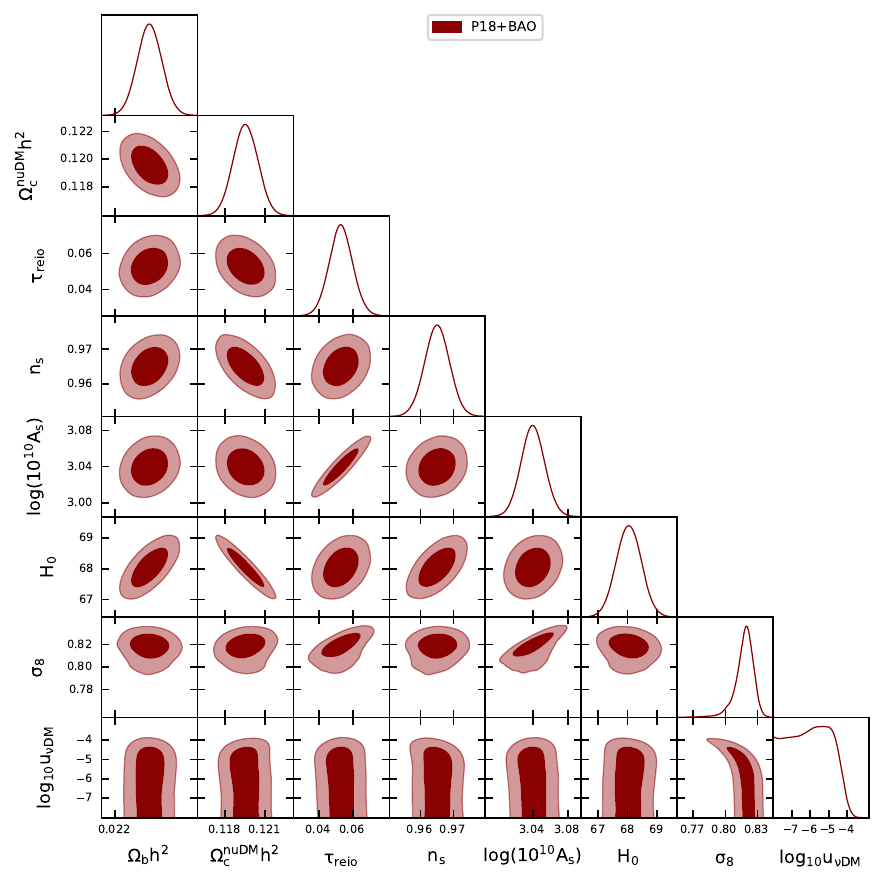}    
\caption{\textbf{Massless neutrinos:} 1-dimensional marginalized posterior distributions and the 2-dimensional joint contours inferred for the most relevant cosmological parameters by analyzing Planck 2018 and BAO data.}\label{fig:nomass_P18}
\end{figure*}

\begin{figure*}[htbp!]  
\includegraphics[width=\textwidth]{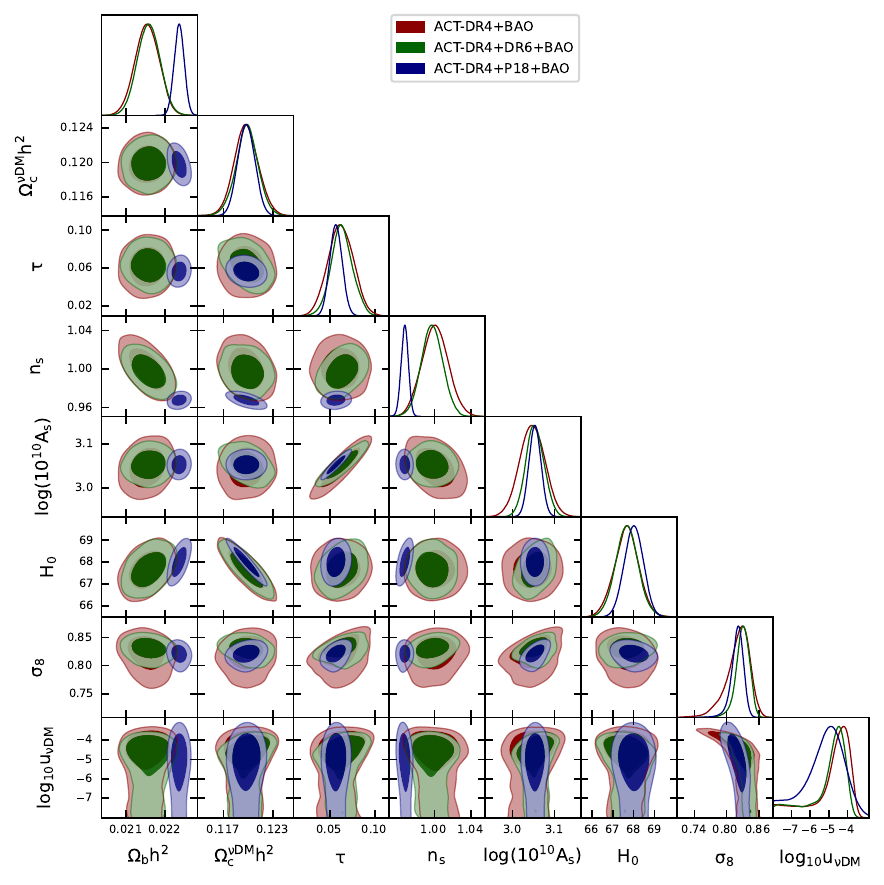}    
\caption{\textbf{Massless neutrinos:} 1-dimensional marginalized posterior distributions and the 2-dimensional joint contours inferred for the most relevant cosmological parameters by analyzing the ACT CMB data and their combinations with Planck and BAO measurements.}
\end{figure*}

\begin{figure*}[htbp!]  
\includegraphics[width=\textwidth]{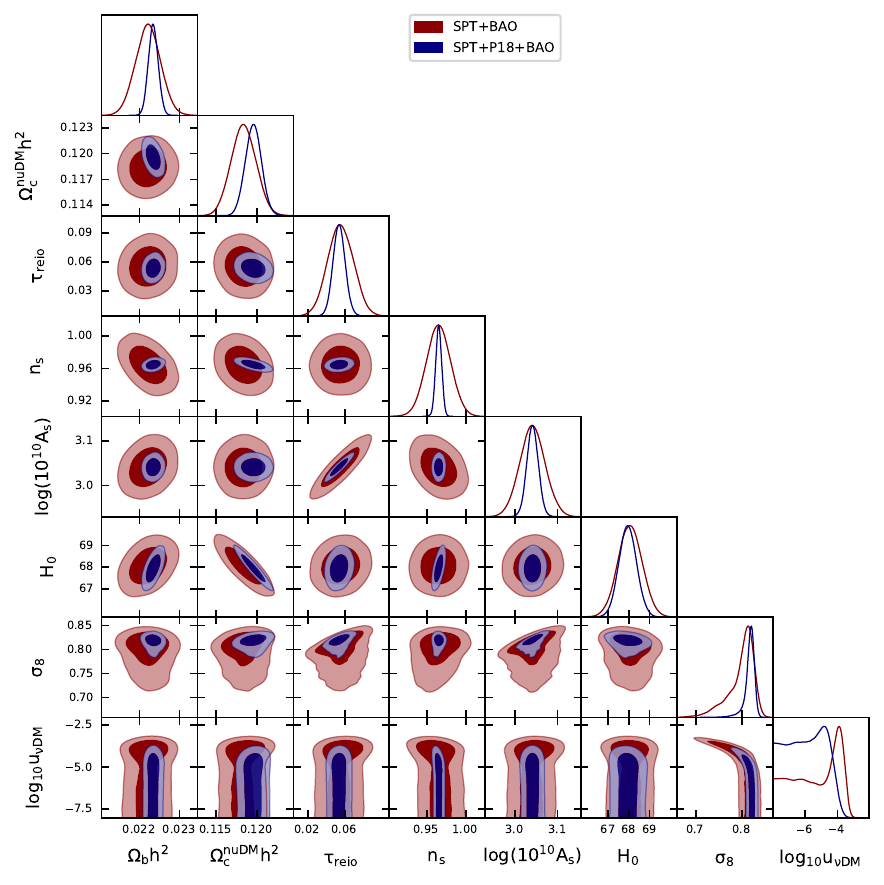}    
\caption{\textbf{Massless neutrinos:} 1-dimensional marginalized posterior distributions and the 2-dimensional joint contours inferred for the most relevant cosmological parameters by analyzing the SPT CMB data and their combinations with Planck and BAO measurements.}
\end{figure*}

\clearpage

\subsection*{Comparison between the Profile and Marginalized Distributions}

\begin{figure*}[htbp!]  
\includegraphics[scale=0.52]{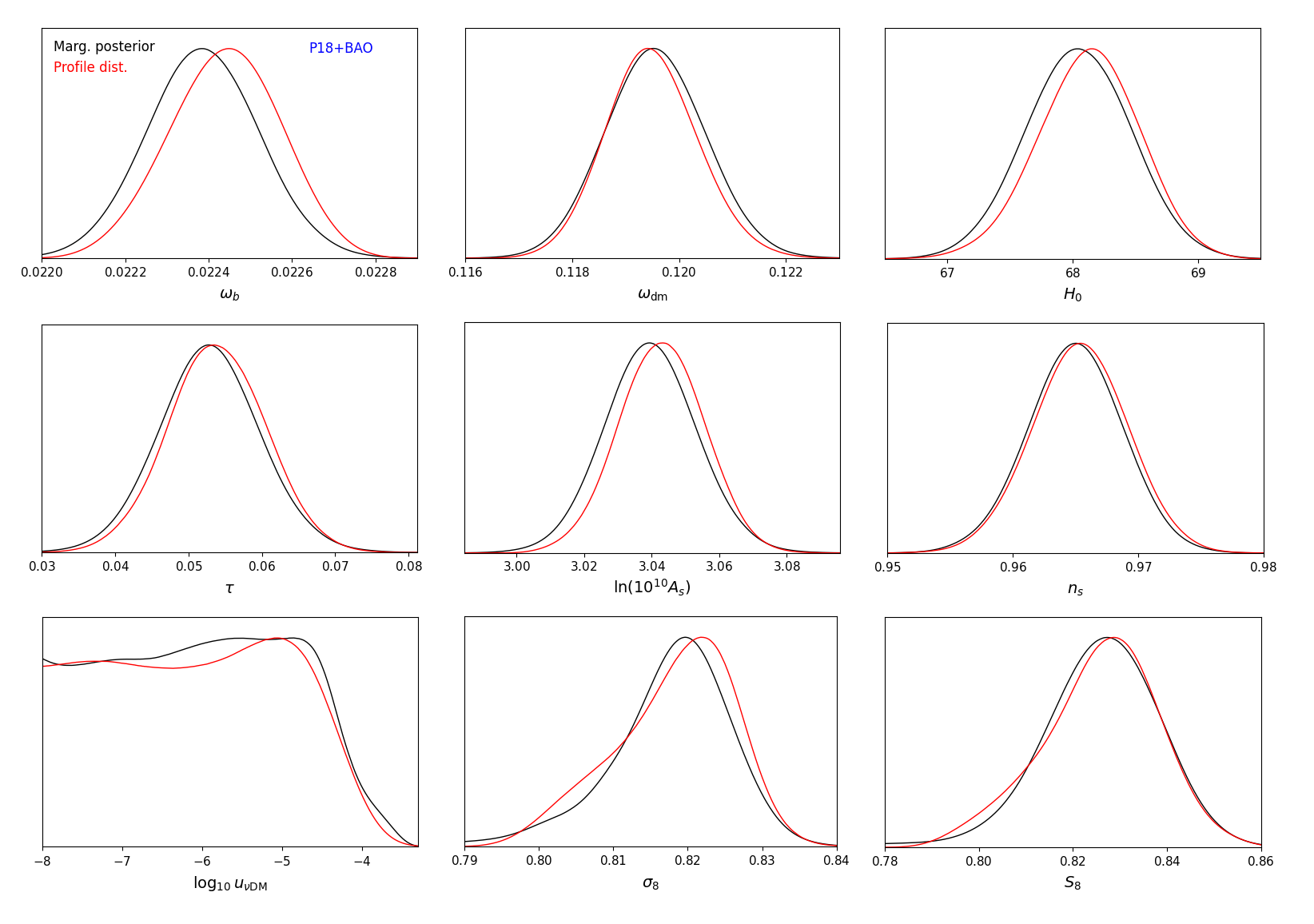}    \caption{\textbf{Massless neutrinos:} One-dimensional marginalized posteriors (in black) and profile distributions (in red) obtained with the P18+BAO data set for the most relevant parameters of the model.}\label{fig:P18+BAO_full}
\end{figure*}

\begin{figure*}[t!]  
\includegraphics[scale=0.52]{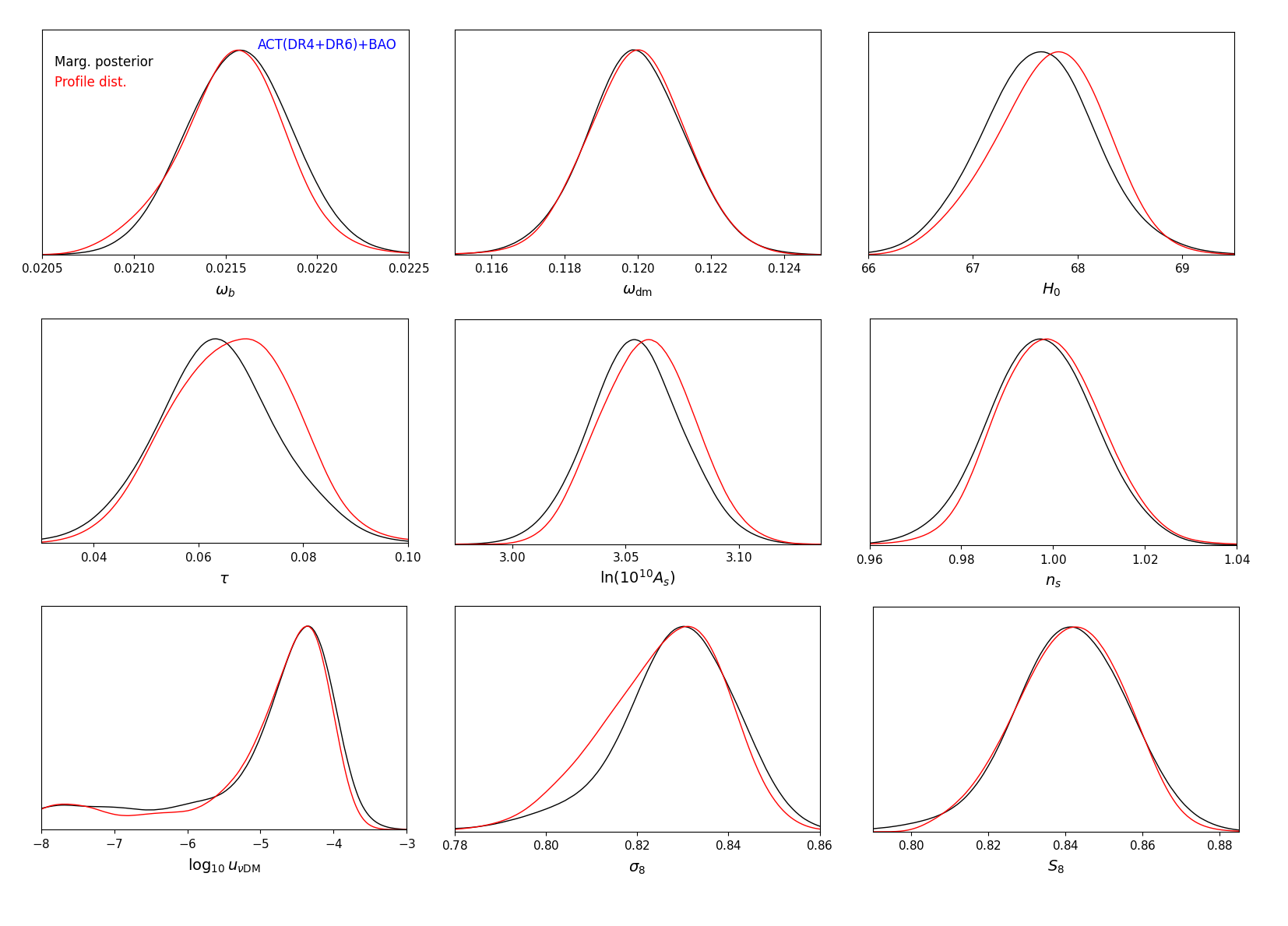}    \caption{\textbf{Massless neutrinos:} As in \autoref{fig:P18+BAO_full}, but employing the ACT (DR4+DR6)+BAO data set.}\label{fig:ACT+lensing+BAO_full}
\end{figure*}

\begin{figure*}[t!]  
\includegraphics[scale=0.52]{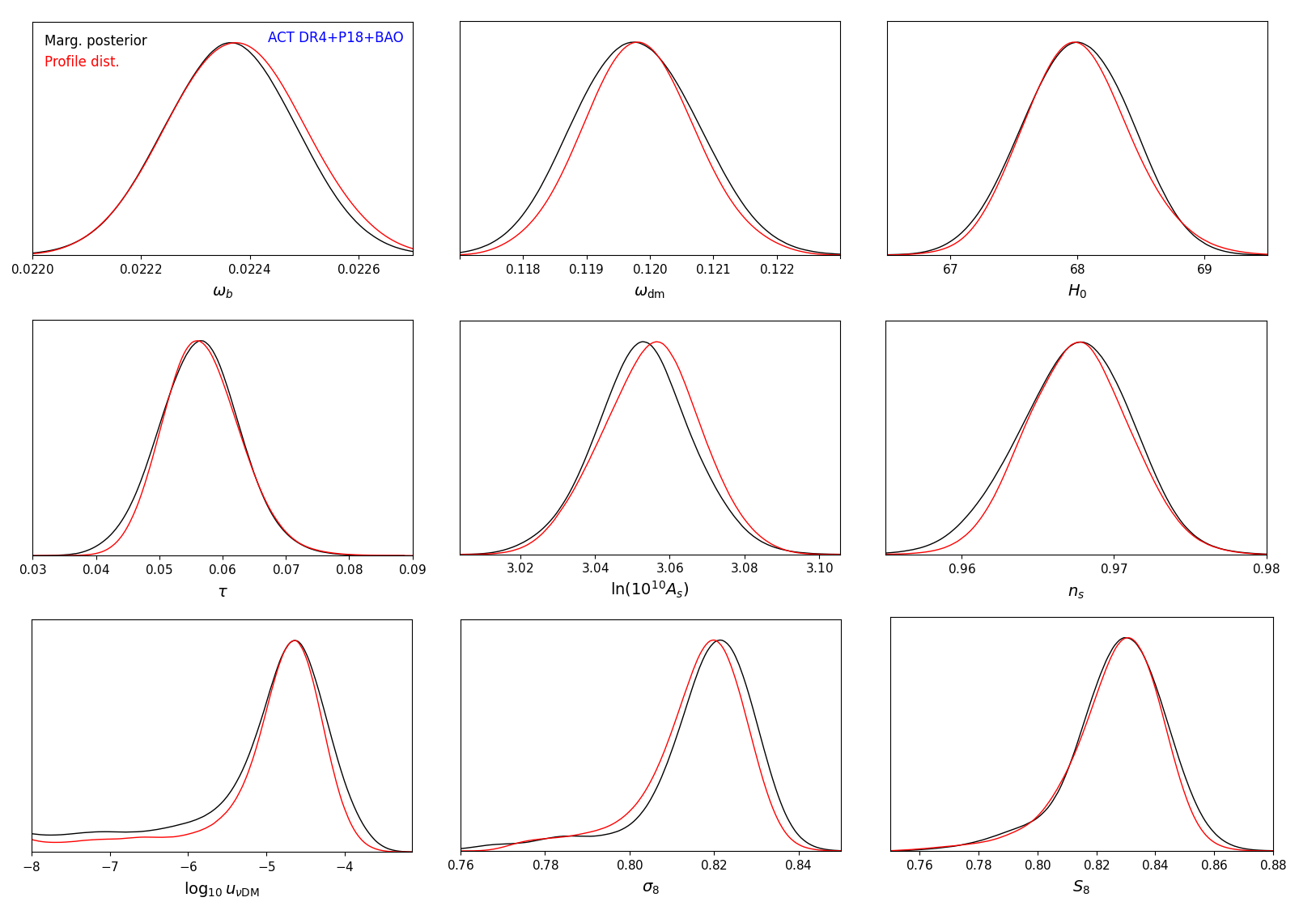}    \caption{\textbf{Massless neutrinos:} As in the previous figures, but using the ACT (DR4)+P18+BAO data set.}\label{fig:ACT+P18+BAO_full}
\end{figure*}

\begin{figure*}[t!]  
\includegraphics[scale=0.52]{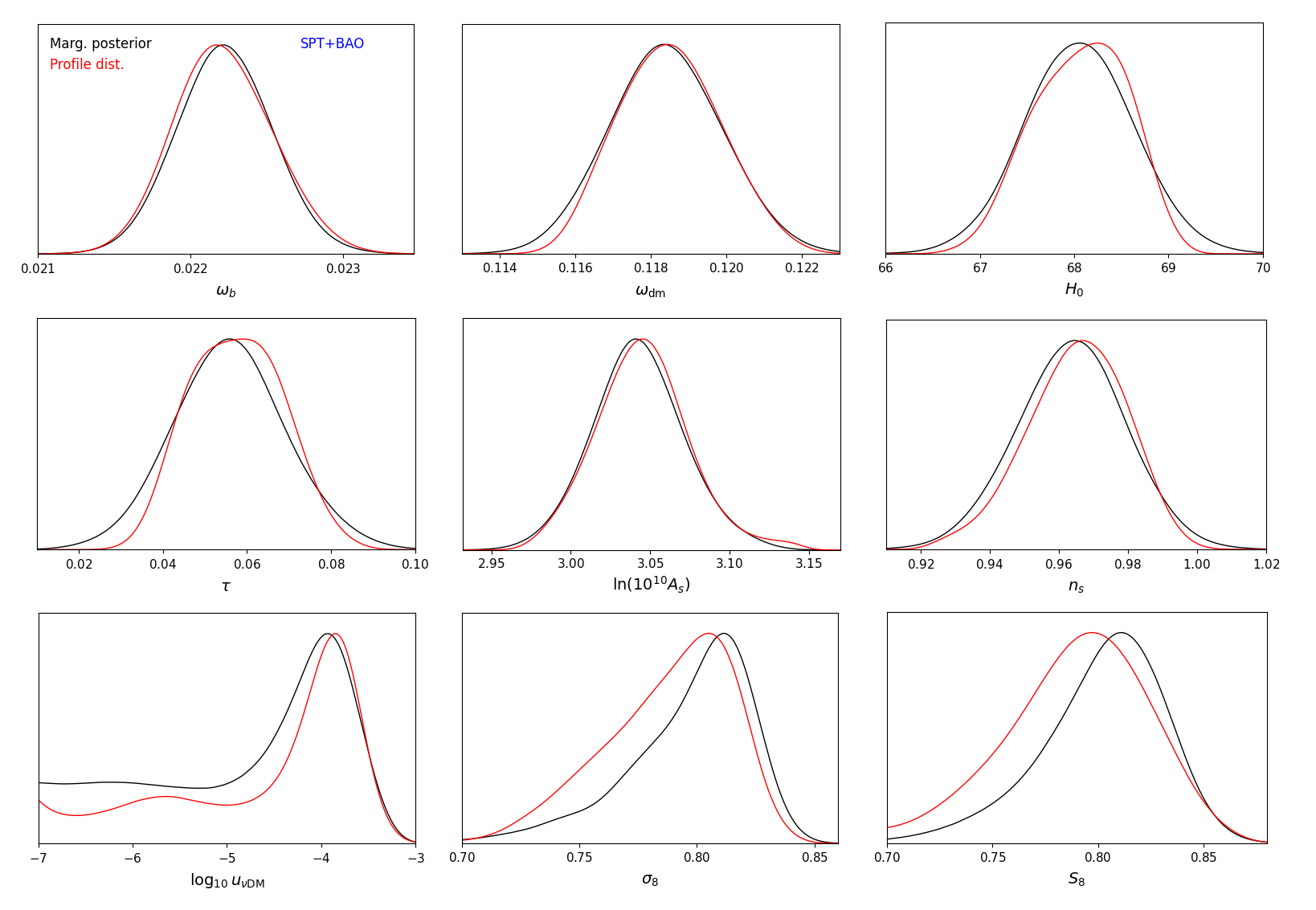}    \caption{\textbf{Massless neutrinos:} As in the previous figures, but making use of the SPT+BAO data set.}\label{fig:SPT+BAO_full}
\end{figure*}

\clearpage

\section*{Massive Neutrinos}

\subsection*{Tables}

\begin{table*}[htbp!] 
\centering
\renewcommand{\arraystretch}{1.5}
\resizebox{\textwidth}{!}{
\begin{tabular}{|c ||c | c || c | c || c | c || c | c || c | c |}
 \multicolumn{1}{c}{} & \multicolumn{2}{c}{\textbf{P18+BAO}} & \multicolumn{2}{c}{\textbf{ACT DR4+P18+BAO}} &\multicolumn{2}{c}{\textbf{ACT (DR4+DR6)+BAO}} &\multicolumn{2}{c}{\textbf{SPT+BAO}} &\multicolumn{2}{c}{\textbf{SPT+P18+BAO}}
\\\hline
{\small Parameter} & {\small Marg.}  & {\small PL} & {\small Marg.}  & {\small PL} & {\small Marg.}  & {\small PL} & {\small Marg.}  & {\small PL} & {\small Marg.}  & {\small PL}
\\\hline
$100\omega_b$ & $2.241\pm 0.013$ & $2.247^{+0.012}_{-0.013}$ & $2.237\pm 0.012$ & $2.239\pm 0.011$  & $2.161\pm 0.030$  & $2.168^{+0.026}_{-0.030}$ & $2.220\pm 0.032$  & $2.223\pm 0.026$ & $2.236\pm 0.012$ & $2.238\pm 0.012$ \\\hline
$10\omega_{\rm dm}$ & $1.192\pm 0.010$ & $1.197^{+0.007}_{-0.010}$ & $1.193\pm 0.010$ & $1.193^{+0.011}_{-0.007}$  & $1.187\pm 0.014$  & $1.192^{+0.012}_{-0.014}$ & $1.159^{+0.020}_{-0.022}$  & $1.164^{+0.016}_{-0.015}$ & $1.191\pm 0.010$ & $1.197^{+0.008}_{-0.011}$\\\hline
$H_0$ & $67.43^{+0.45}_{-0.51}$ & $67.66^{+0.40}_{-0.35}$ & $67.40^{+0.46}_{-0.51}$ & $67.70^{+0.41}_{-0.45}$  & $66.98\pm 0.66$  & $67.43^{+0.48}_{-0.68}$ & $67.47^{+0.63}_{-0.65}$  & $67.91^{+0.40}_{-0.88}$ & $67.37^{+0.47}_{-0.51}$ & $67.55^{+0.49}_{-0.36}$\\\hline
$\tau$ & $0.058\pm 0.007$ & $0.062^{+0.007}_{-0.005}$ & $0.057\pm 0.006$ & $0.057^{+0.007}_{-0.006}$  & $0.075\pm 0.013$  & $0.072^{+0.012}_{-0.011}$ & $0.064\pm 0.014$  & $0.064^{+0.012}_{-0.014}$ & $0.057\pm 0.007$ & $0.055^{+0.007}_{-0.006}$ \\\hline
$n_s$ & $0.966\pm 0.004$ & $0.966^{+0.004}_{-0.003}$ & $0.969\pm 0.004$ & $0.970\pm 0.003$  & $0.996\pm 0.012$  & $0.999\pm 0.011$ & $0.971\pm 0.016$  & $0.985^{+0.008}_{-0.024}$ & $0.966\pm 0.004$ & $0.967\pm 0.003$ \\\hline
$\ln(10^{10}A_s)$& $3.050\pm0.013$  & $3.057^{+0.007}_{-0.018}$   & $3.053^{+0.016}_{-0.012}$   &  $3.053^{+0.018}_{-0.010}$& $3.080\pm 0.024$  & $3.071^{+0.023}_{-0.020}$ & $3.053\pm 0.028$  & $3.062^{+0.023}_{-0.032}$ & $3.046\pm 0.015$ & $3.058^{+0.008}_{-0.014}$\\\hline
$\log_{10}u_{\nu {\rm DM}}$ & $-4.11^{+0.73}_{-0.93}$ & $-5.00^{+0.90}_{-1.80}$ & $-4.19^{+0.39}_{-0.45}$ & $-3.96^{+0.44}_{-0.66}$  & $-4.12^{+0.68}_{-1.32}$  & $-4.00^{+0.59}_{-0.91}$ & $<-3.15$  & $-4.6^{+1.1}_{-1.7}$ & $-5.5\pm 1.2$ & $-5.7\pm 1.2$ \\\hline
$\sum m_{\nu}$ [eV] & $<0.15$ & $<0.15$ & $<0.18$ & $<0.17$  & $<0.24$  & $<0.24$ & $<0.41$  & $<0.37$ & $<0.19$ & $<0.17$ \\\hline
$\sigma_8$ & $0.800^{+0.010}_{-0.011}$ & $0.812^{+0.007}_{-0.011}$ & $0.803^{+0.009}_{-0.013}$ & $0.804^{+0.011}_{-0.009}$  & $0.814\pm 0.013$  & $0.821^{+0.011}_{-0.013}$ & $0.775^{+0.023}_{-0.034}$  & $0.775^{+0.023}_{-0.027}$ & $0.801^{+0.011}_{-0.013}$ & $0.813^{+0.008}_{-0.014}$ \\\hline
$S_8$ & $0.816\pm 0.013$ & $0.824^{+0.009}_{-0.011}$ & $0.819\pm 0.015$ & $0.823^{+0.013}_{-0.012}$  & $0.834\pm 0.015$  & $0.838^{+0.012}_{-0.013}$ & $0.780^{+0.027}_{-0.034}$  & $0.799^{+0.021}_{-0.036}$ & $0.818^{+0.014}_{-0.016}$ & $0.833^{+0.007}_{-0.025}$\\\hline
\end{tabular}
}
\caption{Same as in \autoref{tab:table_masless}, but for massive neutrinos.}\label{tab:table_massive}
\end{table*}

\clearpage

\subsection*{Correlations among cosmological parameters}

\begin{figure*}[htbp!]  
\includegraphics[width=\textwidth]{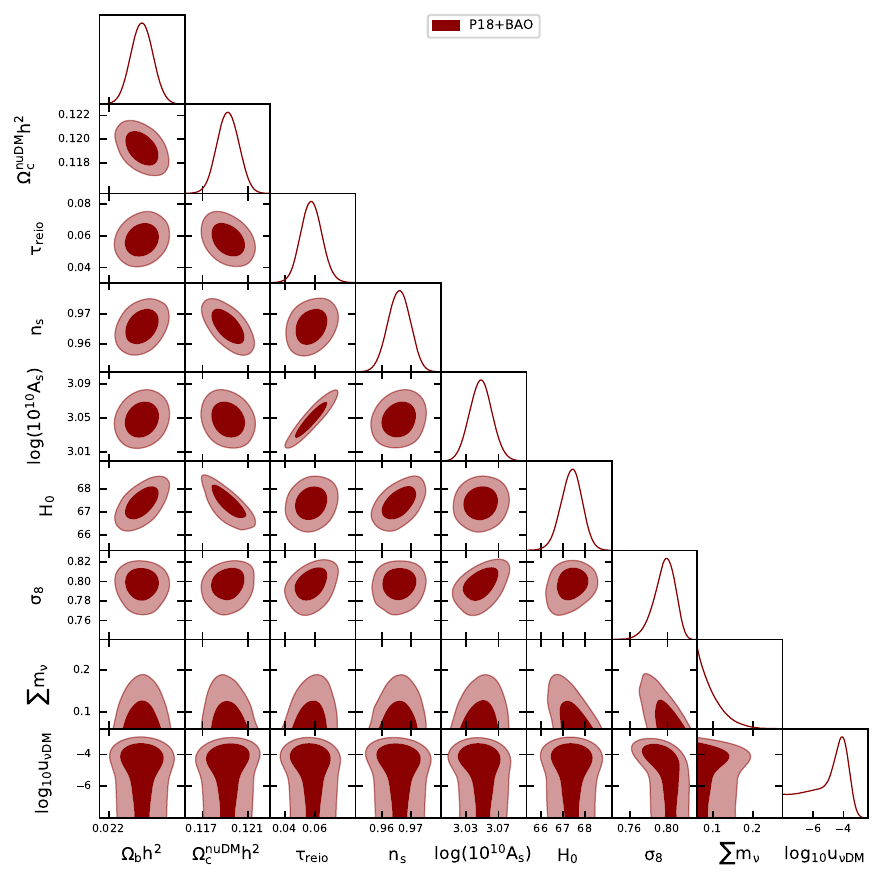}    
\caption{\textbf{Massive neutrinos:} 1-dimensional marginalized posterior distributions and the 2-dimensional joint contours inferred for the most relevant cosmological parameters by analyzing Planck 2018 and BAO data.}\label{fig:P18_massive_contours}
\end{figure*}

\begin{figure*}[htbp!]  
\includegraphics[width=\textwidth]{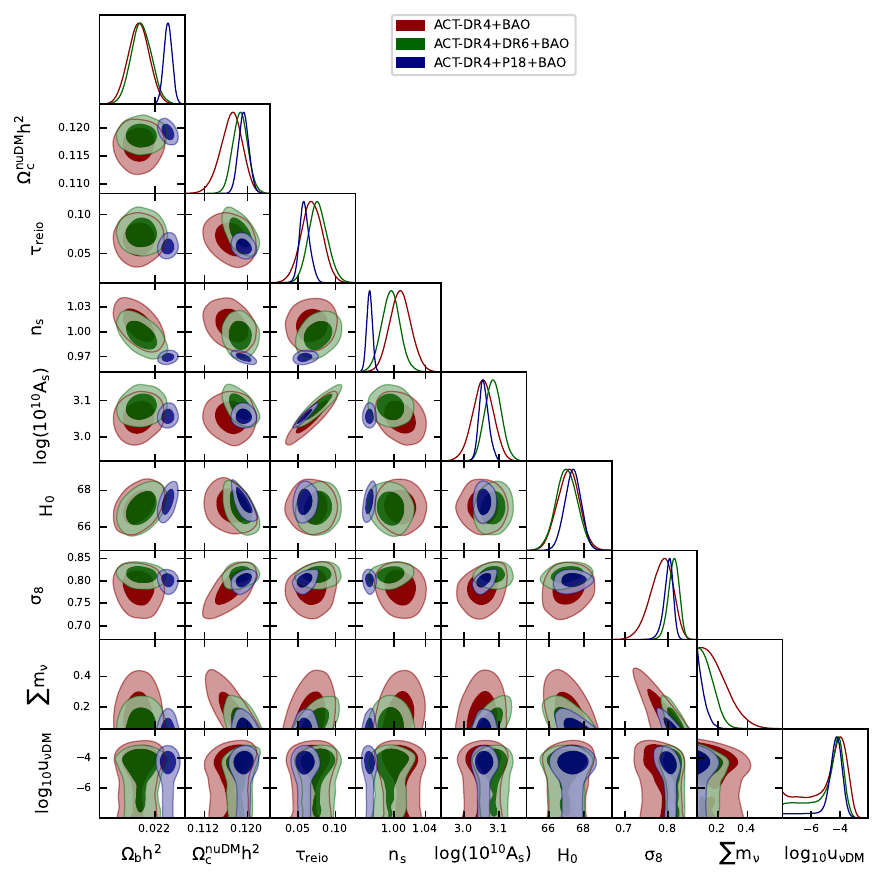}    
\caption{\textbf{Massive neutrinos:} 1-dimensional marginalized posterior distributions and the 2-dimensional joint contours inferred for the most relevant cosmological parameters by analyzing the ACT CMB data and their combinations with Planck and BAO measurements.
}
\end{figure*}

\begin{figure*}[htbp!]  
\includegraphics[width=\textwidth]{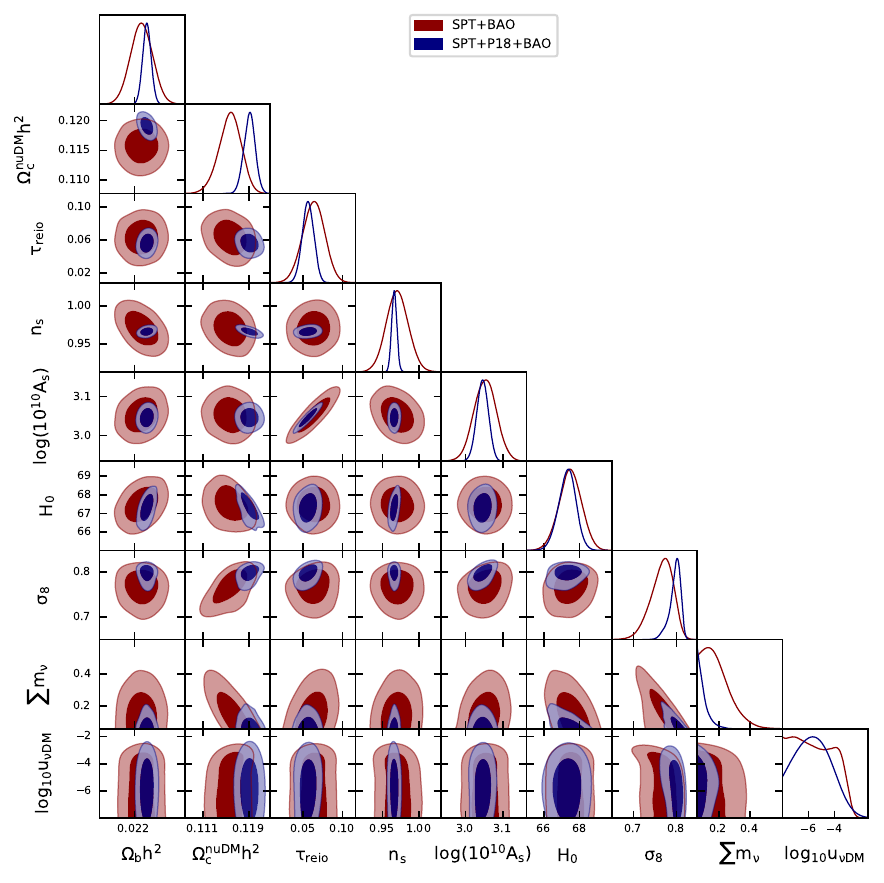}    
\caption{\textbf{Massive neutrinos:} 1-dimensional marginalized posterior distributions and the 2-dimensional joint contours inferred for the most relevant cosmological parameters by analyzing the SPT CMB data and their combinations with Planck and BAO measurements.
}
\end{figure*}

\clearpage

\subsection*{Comparison between the Profile and Marginalized Distributions}

\begin{figure*}[htbp!]  
\includegraphics[scale=0.46]{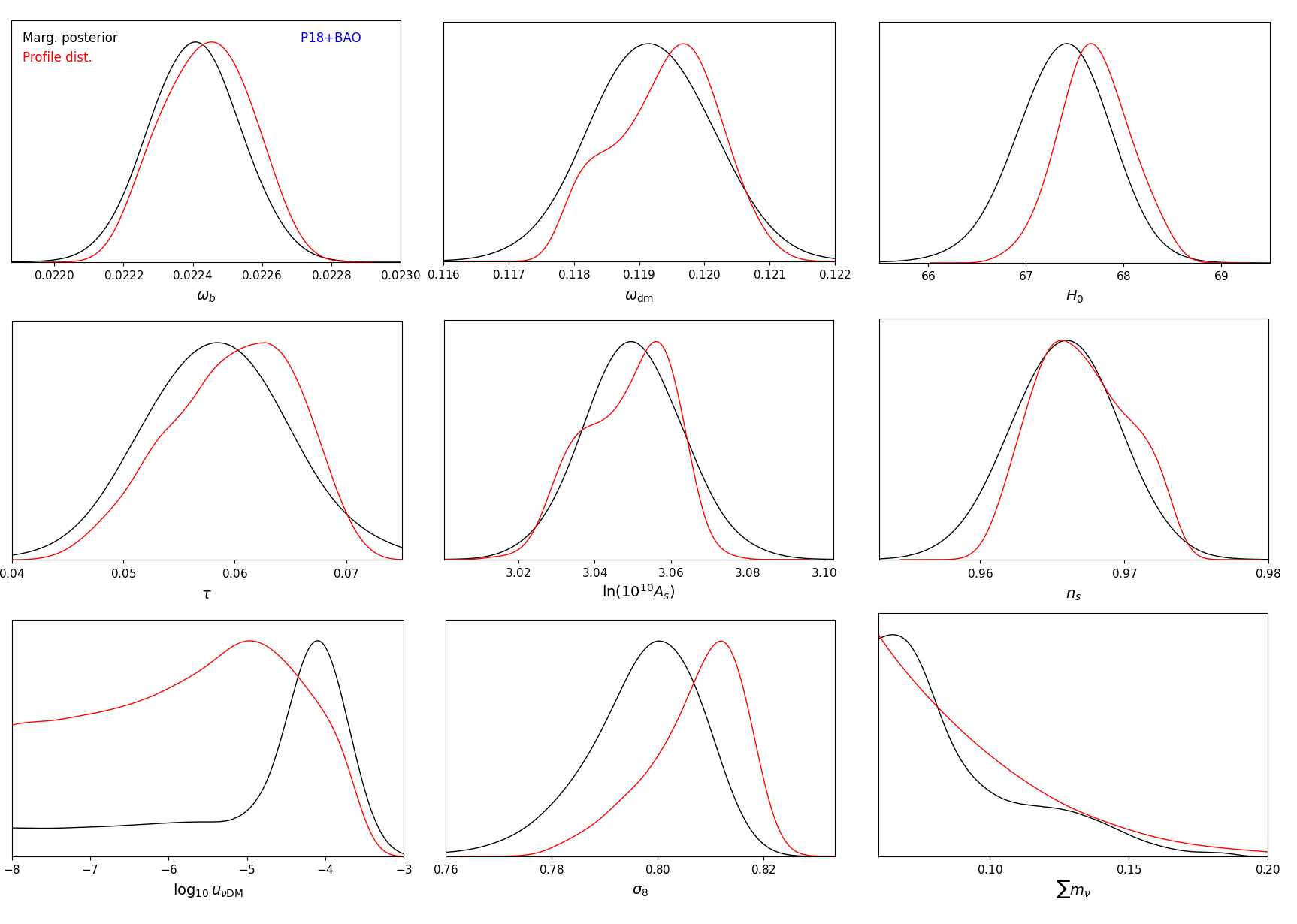}    \caption{\textbf{Massive neutrinos:} One-dimensional marginalized and profile distributions (in black and red, respectively) obtained with the P18+BAO data set and allowing the sum of the neutrino masses to vary freely in the Monte Carlo analysis.}\label{fig:P18+BAO_massive}
\end{figure*}

\begin{figure*}[htbp!]  
\includegraphics[scale=0.52]{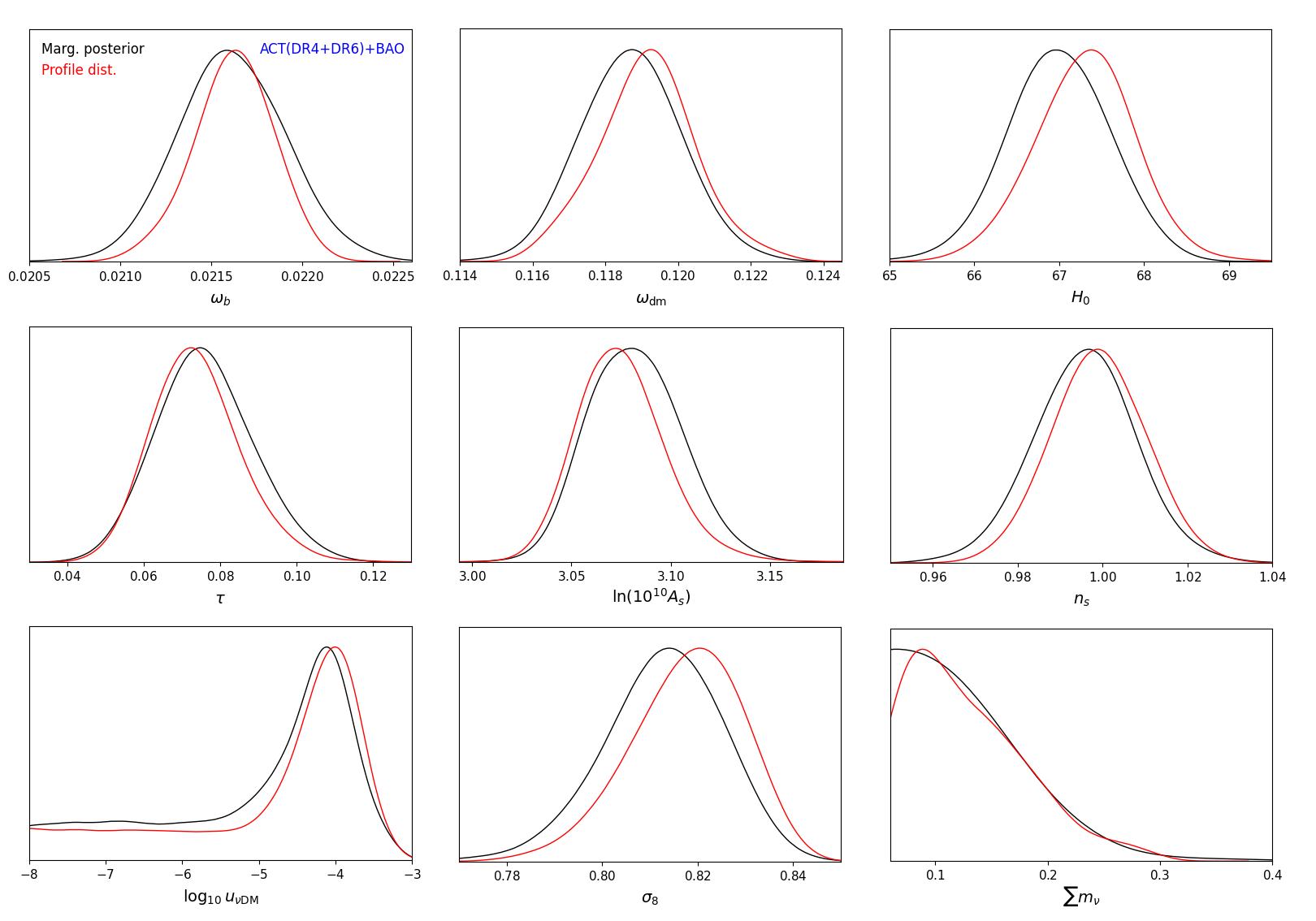}    \caption{ \textbf{Massive neutrinos:} Same as in \autoref{fig:P18+BAO_massive}, but using ACT (DR4+DR6)+BAO.}\label{fig:ACT+lensing+BAO_massive}
\end{figure*}

\begin{figure*}[htbp!]  
\includegraphics[scale=0.52]{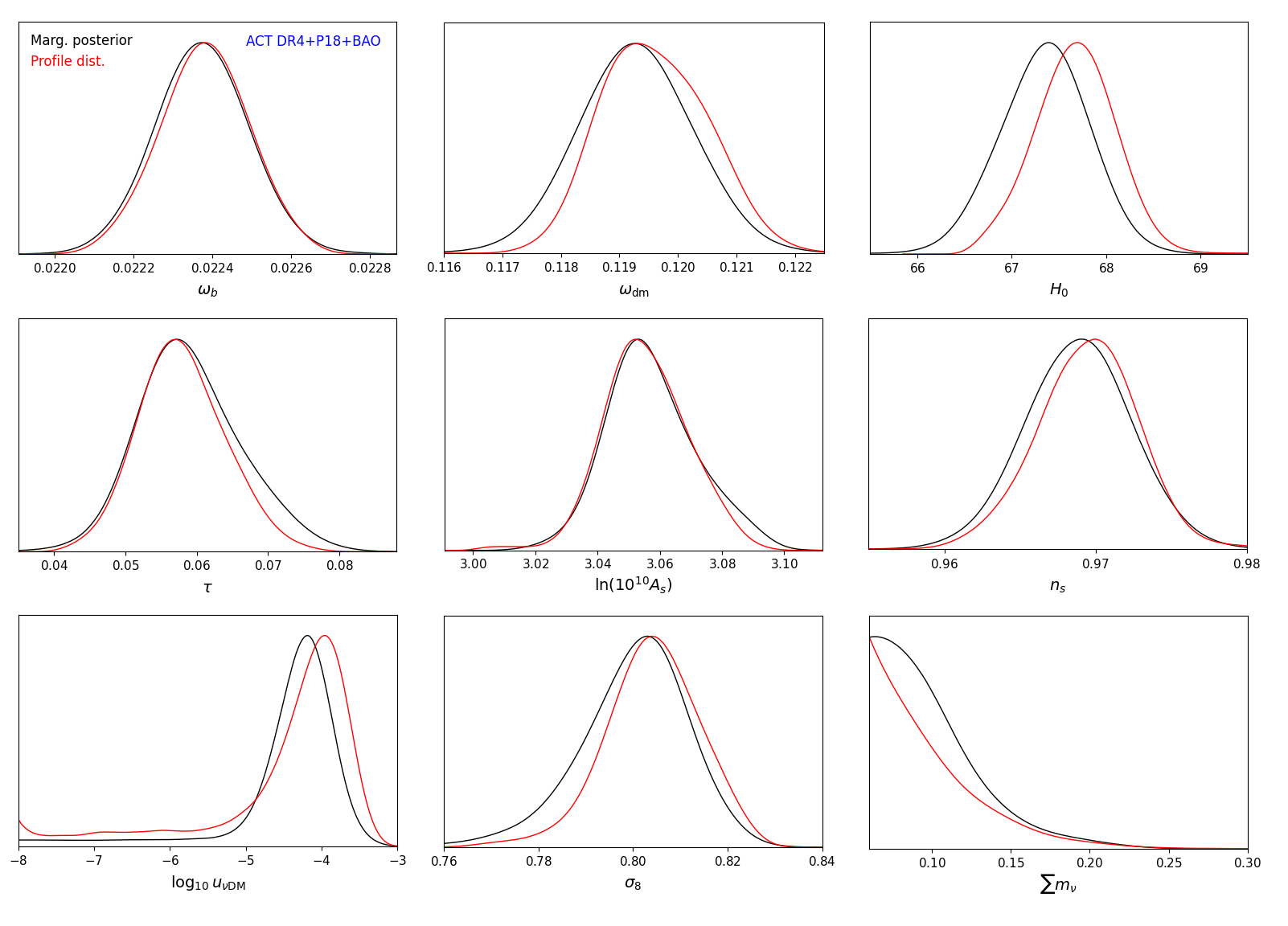}    \caption{\textbf{Massive neutrinos:} Same as in the previous figures, but using ACT DR4+P18+BAO.}\label{fig:ACT+P18+BAO_massive}
\end{figure*}

\begin{figure*}[htbp!]  
\includegraphics[scale=0.52]{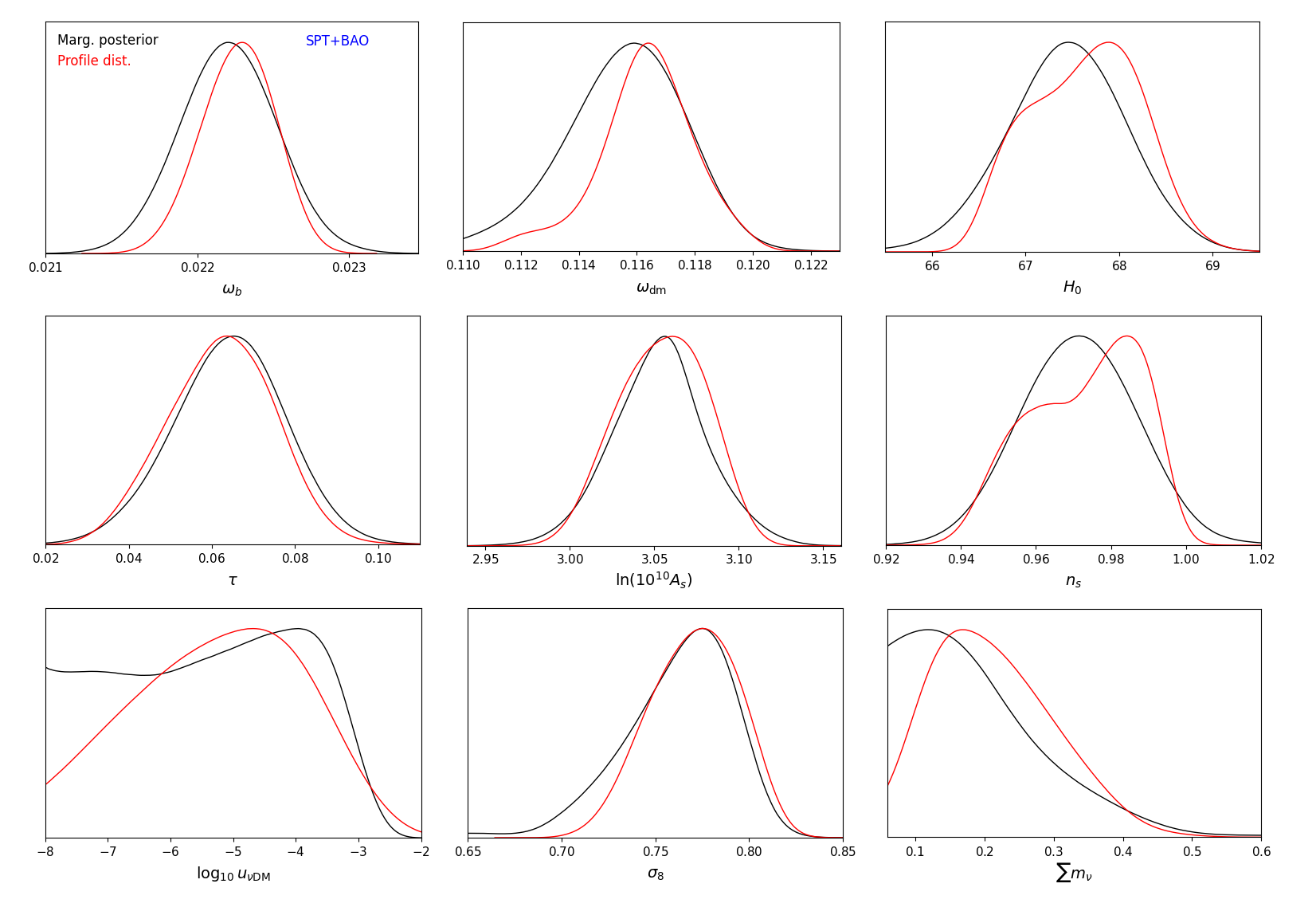}    \caption{ \textbf{Massive neutrinos:} Same as in the previous figures, but using SPT+BAO.}\label{fig:SPT+BAO_massive}
\end{figure*}

\end{document}